\documentclass[aps,prc,reprint,groupedaddress]{revtex4-1} 
\usepackage{amsmath}
\usepackage{graphicx}
\usepackage{array}
\usepackage{capt-of}
\usepackage{comment}
\usepackage[maxfloats=60]{morefloats}
\usepackage[table]{xcolor}
\usepackage{tabulary}
\usepackage{placeins}

\begin{document}
	\newcommand{\PIEL}{$\pi N \rightarrow \pi N$}	
	\newcommand{\GPPN}{$\gamma N \rightarrow \pi N$}
	\newcommand{\GPEP}{$\gamma p \rightarrow \eta p$}
	\newcommand{\GPKL}{$\gamma p \rightarrow K^+ \Lambda$}
	\newcommand{\GNEN}{$\gamma n \rightarrow \eta n$}
	\newcommand{\PINE}{$\pi^- p \rightarrow \eta n$}
	\newcommand{\PINK}{$\pi^- p \rightarrow K^0 \Lambda$}
	\newcommand{\DSG}{$d\sigma / d \Omega$}

\newcommand{\bibE}[4]{#1 \textit{et al.}, #2 \textbf{#3}, #4.}

\newcommand{\pathGPEP}{Figures/GPEP/}
\newcommand{\pathGNEN}{Figures/GNEN/}
\newcommand{\pathGPKL}{Figures/GPKL/}
\newcommand{\pathGPPIP}{Figures/GPPIP/}
\newcommand{\pathGNPIP}{Figures/GNPIP/}
\newcommand{\pathPINE}{Figures/PINE/}
\newcommand{\pathPINK}{Figures/PINK/}

\newcolumntype{C}{@{\extracolsep{0.5cm}}c@{\extracolsep{0pt}}}
\newcommand{\specialcell}[2][c]{\begin{tabular}[#1]{@{}c@{}}#2\end{tabular}}
\renewcommand{\arraystretch}{1.2}

\title{Updated determination of $N^*$ resonance parameters using a unitary, multichannel formalism}


\author{B. C. Hunt and D. M. Manley}
\affiliation{Department of Physics, Kent State University, Kent, OH 44242-0001, USA}


\date{\today}

\begin{abstract}

Results are presented for an updated multichannel energy-dependent partial-wave analysis of $\pi N$ scattering.  Our earlier work incorporated single-energy amplitudes for $\pi N \rightarrow \pi N$, $\gamma N \rightarrow \pi N$, $\pi N \rightarrow \pi\pi N$, $\pi N \rightarrow \eta N$, and $\pi N \rightarrow K \Lambda$.  The present work incorporates new single-energy solutions for $\gamma p \rightarrow \eta p$ up to a c.m.\ energy of 1990~MeV, $\gamma p \rightarrow K^+ \Lambda$ up to a c.m.\ energy of 2230~MeV, and $\gamma n \rightarrow \eta n$ up to a c.m.\ energy of 1885~MeV, as well as updated single-energy solutions for $\pi N \rightarrow \eta N$, $\pi N \rightarrow K \Lambda$, and $\gamma N \rightarrow \pi N$.  In this paper, we present and discuss the resonance parameters obtained from a combined fit of all these single-energy amplitudes.  Our determined energy-dependent amplitudes provide an excellent description of the corresponding measured observables.

\end{abstract}


\maketitle

\section{Introduction}
According to quark models, the baryon is typically viewed as a particle composed of three constituent quarks. With sufficient energy, one or more of the quarks can be excited, giving rise to a spectrum of particles called resonances. The primary experimental method used to search for resonances has been to analyze $\pi N$ reactions including $\pi N \rightarrow \pi N$  and \GPPN. This search has yielded many well-known and established resonances. The first observed resonance, the $P_{33}$(1232), was followed by many others, including the $S_{11}$(1535), $S_{11}$(1650), and $F_{15}$(1680). 

In the literature, there are also many theoretical models \cite{isgur-karl, capstick-isgur, glozman88, loring01, capstick-roberts} that attempt to explain the interactions of the quarks inside of baryons. Each of these theoretical models has one thing in common, they predict more resonances than have been experimentally found. One possible explanation is that these predicted resonances decouple from the $\pi N$ channel. This idea has led to recent experimental efforts using photon beams and meson photoproduction reactions aimed at searching for these resonances. 

To aid in the interpretation of the new data, groups such as EBAC-JLab~\cite{JLABModel}, Bonn-Gatchina~\cite{BnGaModel}, GWU/SAID~\cite{GWUModel}, and KSU~\cite{ManleyModel} have all developed multichannel formalisms to analyze experimental data in a self-consistent framework. The EBAC-JLab group uses a coupled-channel approach that contains the channels $\pi N$, $\pi \pi N$, $\eta N$, $K \Lambda$, and pion photoproduction. Bonn-Gatchina (BnGa) uses a $K$-matrix formalism with Breit-Wigner resonances and includes \PIEL, $\gamma N \rightarrow \pi \pi N$, as well as channels $\eta N$, $K \Lambda$, and $K \Sigma$. The GWU/SAID model is also based on a $K$-matrix approach that focuses on analyses of $\pi N \rightarrow \pi N$~\cite{arndt06} and \GPPN~\cite{Strakovsky15}, but more recent efforts have allowed the inclusion of $\pi N \rightarrow \eta N$ as well.

The KSU model~\cite{ManleyModel} used in this work is based on a generalized energy-dependent Breit-Wigner parametrization of amplitudes that treats all channels on an equal footing, and also takes full account of non-resonant backgrounds. Previous fits using this model included partial-wave amplitudes for  $\pi N \rightarrow \pi N$, $\pi N \rightarrow \pi \pi N$, $\gamma N \rightarrow \pi N$, $\pi N \rightarrow \eta N$, and $\pi N \rightarrow K \Lambda$~\cite{ShresthaED}. The current work updates and supersedes this earlier work by adding single-energy amplitudes for the photoproduction reactions $\gamma p \rightarrow \eta p$  and \GNEN~\cite{OURWORK1} and \GPKL~\cite{OURWORK2}. Our previous single-energy $\pi N \rightarrow \eta N$ and $\pi N \rightarrow K \Lambda$ amplitudes~\cite{shresthaSE} were also updated~\cite{Hunt2017} to be self-consistent with new experimental data for photoproductions reactions having the same final states.

Section~II briefly discusses the formalism behind the generalized $K$-matrix approach. Section~III discusses the fitting procedure used to obtain a fit of the partial-wave amplitudes for each reaction. Section~IV discusses results describing the determined resonance structure. The appendix contains tables of partial widths, branching fractions, and resonant amplitudes. It also contains Argand diagrams showing the final dimensionless energy-dependent partial-wave amplitudes.

\section{Theoretical Model}
In the KSU model~\cite{ManleyModel}, the unitary and symmetric partial-wave scattering matrix $\bf S$, or $S$-matrix, is given by
\begin{equation}
\bf S = B^T R B,
\end{equation}
where $\bf B$ is the product of unitary, symmetric background matrices
\begin{equation}
\bf B = B_1B_2 \cdots B_n
\end{equation}
and $\bf R$ represents the resonant part of the scattering amplitude or $s$-channel process.  Consequently, $\bf B$ itself is unitary but not necessarily symmetric whereas $\bf S$ is both unitary and symmetric.
This is equivalent to the conservation of probability and time-reversal symmetry. The matrix $\bf R$ is constructed by writing
\begin{align}
\begin{split}
\mathbf{R} &= \mathbf{I} + 2\rm{i}\mathbf{T_R} \\
&= \mathbf{I} + 2\rm{i}\mathbf{K}(\mathbf{I}-\rm{i}\mathbf{K})^{-1}\\
&= (\mathbf{I} + \rm{i}\mathbf{K})(\mathbf{I}-\rm{i}\mathbf{K})^{-1},
\label{TvsK}
\end{split}
\end{align}
where $\bf K$ is a Hermitian $K$-matrix, $\mathbf{K} = \mathbf{K}^{\dagger}$. To satisfy time-reversal symmetry, $K$ also must be symmetric. $\bf T_R$ is called the resonant transition matrix, or $T$-matrix for short. Each of the resonances corresponds to a pole in $T_R$ and, therefore, also in the total $S$-matrix. 

In constructing the background, a ``distant poles'' approximation was used. In this approximation, the functional behavior used for the background was a modified Breit-Wigner form where the mass was kept negative and usually large (the majority have magnitudes greater than 3000 MeV, with many larger than $10^4$ MeV). This ensured that the background poles exist far from the physical region of the complex plane. The background terms were also allowed very large widths (on the order of $10^4$~MeV). These features guaranteed that the background had the correct threshold behavior, was slowly varying, and was flexible enough in form to allow the fitting of a large number of potential functional behaviors. 

Because scattering can happen off attractive and repulsive potentials, separate background terms were used for each process. An attractive (repulsive) background was assured by using a positive (negative) width for the background, as explained in~\cite{Niboh97}. In the absence of resonance terms, an attractive (repulsive) background term alone exhibits counter-clockwise (clockwise) motion on an Argand diagram, but such background amplitudes (unlike resonant amplitudes) do not cross the imaginary axis.

All amplitudes used in the parametrization are dimensionless by construction, while the single-energy photoproduction amplitudes \cite{OURWORK1, OURWORK2} have dimensions of mfm. Once an initial single-energy fit has been performed, the dimensioned single-energy amplitudes are converted to dimensionless amplitudes using~\cite{Niboh97}:

\begin{subequations}
\begin{align}
\begin{split}
\tilde{E}_{l+}& = C_I\sqrt{kq\left(l+1\right)\left(l+2\right)}\: E_{l+},
\end{split}\\[3pt]
\begin{split}
\tilde{E}_{\left(l+1\right)-}& = C_I\sqrt{kql\left(l+1\right)} \: E_{(l+1)-},
\end{split}\\[3pt]
\begin{split}
\tilde{M}_{l+}  &= C_I\sqrt{kql\left(l+1\right)} \: M_{l+}, 
\end{split}\\[3pt]
\intertext{and}
\begin{split} 
\tilde{M}_{\left(l+1\right)-}& = C_I\sqrt{kq\left(l+1\right)\left(l+2\right)} \:      M_{(l+1)-},
\end{split}
\end{align}
\label{Conversion1}

\end{subequations}
\noindent
where the multipoles with a tilde denote the dimensionless amplitudes. Here $C_I$ is an isospin coefficient.  For $\gamma N \rightarrow \eta N$ and $\gamma N \rightarrow K \Lambda$, $C_{1/2} = 1$ and $C_{3/2} = 0$. For $\gamma N \rightarrow \pi N$, $C_{1/2} = -\sqrt{3}$ and $C_{3/2} = \sqrt{2/3}$.  For $\gamma N \rightarrow \pi N$, $k$ and $q$ are the c.m.\ momentum for the incoming $\gamma N$ and outgoing $\pi N$, respectively, and similarly for $\gamma N \rightarrow \eta N$ and $\gamma N \rightarrow K \Lambda$.

The model contains resonance and background couplings for the reactions $\pi N \rightarrow \pi N$, $\pi N \rightarrow \pi \pi N$, $\pi N \rightarrow \eta N$, $\pi N \rightarrow K \Lambda$, \GPEP, \GNEN, and \GPPN \ all of which have single-energy amplitudes determined. It also includes channels that have not been analyzed to date (such as $\rho \Delta $, $\omega N$, and $\pi N^*$), which are included in fits as ``dummy channels'' to satisfy unitarity and prevent over-saturating couplings for measured channels.

\section{Fitting Procedure}

The fitting procedure for obtaining resonance parameters consisted of a two-step process. The first step was to determine single-energy partial-wave amplitudes independent of any resonance structure by fitting observables data in specified energy bins. The single-energy amplitudes for a given partial wave (e.g., $S_{11}$ or $P_{11}$) were then fitted as real and imaginary parts with our energy-dependent parametrization to update the resonance parameters and determine corresponding energy-dependent amplitudes. This procedure was iterated until the energy-dependent solution provided a good description of the observables data. The procedure used for fitting was the standard $\chi^2$ minimization technique. 

To gain confidence in both model stability and reaching a global $\chi^2$ minimum, two techniques were used. The first was to start from a number of distinct solutions and test for convergence in the solution. For this procedure, a local minimum for each starting point was found using the two-step convergence procedure. Each minimum could then be compared to other local minima previously obtained for both a single reaction as well as for all combined reactions. An optimal solution is then one that is sufficiently close to a global minimum for each individual reaction as well as for all reactions combined. The second technique was a randomization process that was devised as follows. A group of resonance parameters were selected to be randomly varied, with each parameter's random variation independently determined and small. (For instance, the parameters might be all couplings to all $P_{11}$ resonances.) The random change for the parameters was kept small, usually less than 20$\%$ of their starting values. By performing these techniques hundreds of times on different subsets of parameters over the course of the analysis, a large region of parameter space was analyzed and checked. This technique also led to confidence that the determined error bars were reasonable. 

To determine final error bars for the single-energy amplitudes, the moduli for each of the partial-wave amplitudes over all newly added photoproduction reactions were treated as free parameters and allowed to vary one final time while the phases were kept fixed in a ``zero-iteration'' fit. This is described in greater detail in the papers describing the single-energy analysis~\cite{OURWORK1, OURWORK2}. The next step was to put these single-energy amplitudes with their final error bars into the energy-dependent code to generate final error bars for all resonance parameters. In this fit, parameter values were not actually varied and the only purpose of the ``fit'' was to calculate error bars taking into account all the various correlations between free parameters. The single-energy points that generated a large contribution to $\chi^2$ had their error bars scaled up until the $\chi^2$ contribution from those points equaled four. This scaling was done to keep individual points from dominating the results for the fits. Then a full error matrix was calculated with a zero-iteration fit to give the final error bars with all parameters treated as free parameters, but not actually varied. Lastly, the uncertainties in the resonance parameters were scaled by $\sqrt{\chi^2/\nu}$, where $\nu$ was the number of degrees of freedom for the fit.

\section{Results}
This section is laid out as follows. Section \ref{Iso12Results} contains information about each of the isospin-1/2 amplitudes and tables of their respective resonance parameters and helicity couplings. Section \ref{Iso32Results} contains information about the isospin-3/2 amplitudes and tables of their respective resonance parameters and helicity couplings.
\subsection{Results for Isospin-1/2 Amplitudes}
\label{Iso12Results}
The following section discusses results for the isospin-1/2 amplitudes. Tables \ref{MassPole1} and \ref{MassPole2} contain the masses ($M$), widths ($\Gamma$), pole positions, and branching fractions of each isospin-1/2 resonance with errors on the last reported significant figure shown in parentheses. Only masses are quoted for resonances above 2300~MeV because their widths and couplings are not reliable at this stage of analysis. Tables \ref{helicity12_SPD} and \ref{Helicity12_DFG} show helicity couplings to the isospin-1/2 resonances. Comparisons are made in each table with  ~\cite{julich15}, ~\cite{anisovich12}, and ~\cite{workman12}. Additional comparisons can be found in the {\textit{Review of Particle Physics}} (RPP)~\cite{pdg}. Partial widths, branching fractions, and resonant amplitudes ($\sqrt{xx_i}$) are listed in Tables \ref{PartialWidths1} and \ref{PartialWidths2} of the appendix. Finally, the energy-dependent fits for each reaction and resonance are shown in Figs.~\ref{Argand01} - \ref{Argand26} of the appendix. 

\begin{table*}[hbtp]
	
	\begin{tabular}{@{\extracolsep{1pt}}cccclCcccl@{}}
		\hline \hline
		\raisebox{-0.5ex}{~~\specialcell {Mass \\(MeV)}~~}& \raisebox{-0.5ex}{~~\specialcell {Width \\(MeV)}~~} & \raisebox{-0.5ex}{~~\specialcell {Re Pole \\(MeV)}~~} & \raisebox{-0.5ex}{\specialcell {$-2$ Im Pole \\(MeV)}} &
		\raisebox{-0.5ex}{\bf ~Analysis} &
		\raisebox{-0.5ex}{~~~\specialcell {Mass \\(MeV)}~~~}& \raisebox{-0.5ex}{~~\specialcell {Width \\(MeV)}~~} & \raisebox{-0.5ex}{~~\specialcell {Re Pole \\(MeV)}~~} &
		\raisebox{-0.5ex}{\specialcell {$-2$ Im Pole \\(MeV)}} &\raisebox{-0.5ex}{\bf ~Analysis}\\[1.1ex] \hline			
		\multicolumn{5}{c}{\raisebox{-2.0ex}{\large \bf $S_{11}(1535)$****}} & \multicolumn{5}{c}{\raisebox{-2.0ex}{\large \bf $S_{11}(1650)$****}}\\[2.5ex]
		\cline{1-5} \cline{6-10}
		
		$1525(2)$        & $147(5)$         & $1496$            & $119$      & This~Work  & $1525(2)$        & $147(5)$         & $1496$            & $119$      & This~Work\\
		                 &                  & $1499$            & $104$      & R\"onchen 15B   &             &                   & $1672$            & $137$      & R\"onchen 15B\\
		$1547$           &$188(14)$         &                   &            & Workman 12   & $1635$           & $115(14)$        &                   &            & Workman 12\\
		$1519(5)$        &$128(14)$         & $1501(4)$         &  $134(11)$  & Anisovich 12   &   $1651(6)$        & $104(10)$        & $1647(6)$         &  $103(8)$  & Anisovich 12\\
		\hline 
		\multicolumn{5}{c}{\raisebox{-2.0ex}{\large \bf $S_{11}(1895)$****}} & \multicolumn{5}{c}{\raisebox{-2.0ex}{\large \bf $P_{11}(1440)$****}}\\[2.5ex]
		\cline{1-5} \cline{6-10}
		$2000(29)$        & $466(72)$         & $1956$            & $449$      & This~Work  & $1417(4)$        & $257(11)$         & $1360$            & $186$      & This~Work\\
		--     &      --          &     --        &   --       & R\"onchen 15B   &     --        &    --           & $1355$            & $215$      & R\"onchen 15B\\
			--     &      --          &     --        &   --       & Workman 12   & $1485(1)$        & $284(4)$         &      --        &     --     & Workman 12\\
		$1895(15)$       & $90^{+30}_{-15}$ & $1900(15)$        &  $90^{+30}_{-15}$ & Anisovich 12   & $1430(8)$        &$365(35)$         & $1370(4)$         &  $190(7)$  & Anisovich 12 \\
		\hline 
		\multicolumn{5}{c}{\raisebox{-2.0ex}{\large \bf $P_{11}(1710)$****}} & \multicolumn{5}{c}{\raisebox{-2.0ex}{\large \bf $P_{11}(1880)$***}}\\[2.5ex]
		\cline{1-5} \cline{6-10}
		$1648(16)$        & $195(46)$         & $1615$            & $169$      & This~Work  & $1967(20)$       & $500(77)$        & $1880$            & $429$      & This~Work\\
		--       &      --        & $1651$    & $121$      & R\"onchen 15B    &         --         &      --            & $1747$            & $323$      & R\"onchen 15B\\
		--     &       --           &        --           &     --       & Workman 12   & --     &        --          &     --         &   --         & Workman 12\\
		$1710(20)$       &$200(18)$         & $1687(17)$        &  $200(25)$ & Anisovich 12   & $1870(35)$       & $235(65)$        & $1860(35)$        &  $250(70)$ & Anisovich 12 \\
		\hline 
		\multicolumn{5}{c}{\raisebox{-2.0ex}{\large \bf $P_{11}(2100)$***}} & \multicolumn{5}{c}{\raisebox{-2.0ex}{\large \bf $P_{13}(1720)$****}}\\[2.5ex]
		\cline{1-5} \cline{6-10}
		$2221(92)$        & $545(170)$         & $2217$            & $545$      & This~Work  & $1711(4)$        & $229(22)$         & $1654$            & $100$      & This~Work\\
        	--	&      --       & --            & --    & R\"onchen 15B   &   --       &   --      & $1710$            & $219$      & R\"onchen 15B\\
		--           & --         &      --          &   --         & Workman 12   & $1764$           & $210$            &   --          &    --        & Workman 12\\
		--        & --       & --         &  --  & Anisovich 12   & $1690^{+70}_{-35}$ &$420(100)$      & $1660(30)$        & $450(100)$ & Anisovich 12 \\
		\hline 
		\multicolumn{5}{c}{\raisebox{-2.0ex}{\large \bf $P_{13}(1900)$****}} & \multicolumn{5}{c}{\raisebox{-2.0ex}{\large \bf $P_{13}(2040)$*}}\\[2.5ex]
		\cline{1-5} \cline{6-10} 
		$1911(6)$        & $292(16)$         & $1856$            & $241$      & This~Work  & $2244(30)$        & $530(89)$         & $2231$            & $529$      & This~Work\\
		--     &    --         &     --         &     --    & R\"onchen 15B  & --        & --        & --            & --      & R\"onchen 15B\\
		--     &    --         &     --         &     --    & Workman 12   & --        & --         & --            & --     & Workman 12 \\
		$1905(30)$       & $250^{+120}_{-50}$ & $1900(30)$      &  $200^{+100}_{-60}$ & Anisovich 12   & $1525(2)$        & $147(5)$         & $1496$            & $119$      & R\"onchen 15B \\ \hline	
		\multicolumn{5}{c}{\raisebox{-2.0ex}{\large \bf $D_{13}(1520)$****}} & \multicolumn{5}{c}{\raisebox{-2.0ex}{\large \bf $D_{13}(1700)$***}}\\[2.5ex]
		\cline{1-5} \cline{6-10}
		$1512.0(1.5)$    & $121(3)$         & $1500$            & $117$      & This~Work  & $1653(5)$        & $81(13)$          & $1647$            & $79$       & This~Work\\
		  --             &   --             & $1512$            & $89$       & R\"onchen 15B   & --     &    --            &    --             &     --     & R\"onchen 15B\\
		$1515$           & $104$            &   --              &   --       & Workman 12   & --     &    --            &    --             &     --     & Workman 12\\
		$1517(3)$        & $114(5)$         & $1507(3)$         & $111(5)$   & Anisovich 12   & $1790(40)$       & $390(140)$       & $1770(40)$        & $420(180)$ & Anisovich 12 \\ \hline
		\multicolumn{5}{c}{\raisebox{-2.0ex}{\large \bf $D_{13}(1875)$***}} & \multicolumn{5}{c}{\raisebox{-2.0ex}{\large \bf $D_{13}(2120)$***}}\\[2.5ex]
		\cline{1-5} \cline{6-10}
		$2005(12)$       & $321(21)$         & $1993$            & $319$      & This~Work  & $2353(29)$       & $503(62)$        & $2357$            & $503$      & This~Work\\
		--     &    --            &    --             &     --     & R\"onchen 15B   & --     &    --            &    --             &     --     & R\"onchen 15B\\
		--     &    --            &    --             &     --     & Workman 12   & --     &    --            &    --             &     --     & Workman 12\\
		$1880(20)$    & $200(25)$          & $1860(25)$         & $200(20)$  & Anisovich 12   & $2150(60)$       & $330(45)$        & $2110(50)$        & $340(45)$  & Anisovich 12 \\ \hline	\hline										
	\end{tabular}
	
	\caption{\label{MassPole1} Comparison of $S_{11}$, $P_{11}$, $P_{13}$, and $D_{13}$ resonance masses, widths, and pole positions for isospin-1/2 amplitudes. Star rating is that found in the RPP~\cite{PDG2018}. Comparisons are made with works by R\"onchen 15b~\cite{julich15}, Anisovich 12~\cite{anisovich12}, and SAID~\cite{workman12}.}
\end{table*}

\begin{table*}[hbtp]	
	\begin{tabular}{@{\extracolsep{1pt}}cccclCcccl@{}}
		\hline \hline
		\raisebox{-0.5ex}{~~\specialcell {Mass \\(MeV)}~~}& \raisebox{-0.5ex}{~~\specialcell {Width \\(MeV)}~~} & \raisebox{-0.5ex}{~~\specialcell {Re Pole \\(MeV)}~~} & \raisebox{-0.5ex}{\specialcell {$-2$ Im Pole \\(MeV)}} &
		\raisebox{-0.5ex}{\bf ~Analysis} &
		\raisebox{-0.5ex}{~~~\specialcell {Mass \\(MeV)}~~~}& \raisebox{-0.5ex}{~~\specialcell {Width \\(MeV)}~~} & \raisebox{-0.5ex}{~~\specialcell {Re Pole \\(MeV)}~~} &
		\raisebox{-0.5ex}{\specialcell {$-2$ Im Pole \\(MeV)}} &\raisebox{-0.5ex}{\bf ~Analysis}\\[1.1ex] \hline			
		\multicolumn{5}{c}{\raisebox{-2.0ex}{\large \bf $D_{15}(1675)$****}} & \multicolumn{5}{c}{\raisebox{-2.0ex}{\large \bf $D_{15}(2060)$***}}\\[2.5ex]
		\cline{1-5} \cline{6-10}
		
		$1669(2)$        & $161(8)$         & $1646$            & $146$        & This~Work  & $2111(17)$        & $499(70)$         & $2010$            & $395$      & This~Work\\
		 --              &   --             & $1646$            & $125$        & R\"onchen 15B   & --     &     --         &     --        &  --      & R\"onchen 15B\\
		$1674(1)$        & $147$            &  --               &  --           & Workman 12   & --     &     --         &     --        &  --      & Workman 12\\
		$1519(5)$        &$128(14)$         & $1501(4)$         &  $134(11)$  & Anisovich 12   & $2060(15)$       & $375(25)$         & $2040(15)$       & $390(25)$  & Anisovich 12 \\
		\hline 
		\multicolumn{5}{c}{\raisebox{-2.0ex}{\large \bf $F_{15}(1680)$****}} & \multicolumn{5}{c}{\raisebox{-2.0ex}{\large \bf $F_{15}(1860)$**}}\\[2.5ex]
		\cline{1-5} \cline{6-10}
		$1681.0(1)$      & $123(3)$         & $1668$            & $118$      & This~Work    & $1928(21)$       & $376(58)$        & $1871$            & $337$      & This~Work\\
		   --            &   --             & $1669$            & $100$      & R\"onchen 15B  &  --     &     --         &     --        &  --      & R\"onchen 15B\\
		$1680$           & $128$            &   --              &   --       & Workman 12   & --     &     --         &     --        &  --      & Workman 12\\
		$1689(6)$        & $118(6)$         & $1676(6)$         &  $113(4)$  & Anisovich 12   & $1860_{-60}^{+120}$ & $270_{-50}^{+140}$ & $1830_{-60}^{+120}$       &  $250^{+150}_{-50}$      & Anisovich 12 \\
		\hline 
		\multicolumn{5}{c}{\raisebox{-2.0ex}{\large \bf $F_{17}(1990)$**}} & \multicolumn{5}{c}{\raisebox{-2.0ex}{\large \bf $F_{17}(2200)$ new}}\\[2.5ex]
		\cline{1-5} \cline{6-10}
		$2028(19)$       & $490(110)$       & $1913$            & $163$      & This~Work  & $2219(16)$        & $519(94)$         & $2106$            & $385$      & This~Work\\
		--               &  --              & $1738$            & $188$      & R\"onchen 15B   & --     &     --         &     --        &  --      & R\"onchen 15B\\
		--               & --               &   --              &   --       & Workman 12   & --     &     --         &     --        &  --      & Workman 12\\
		$2060(65)$       & $240(50)$        & $2030(65)$        & $240(60)$  & Anisovich 12   & --     &     --         &     --        &  --      & Anisovich 12 \\
		\hline 
		\multicolumn{5}{c}{\raisebox{-2.0ex}{\large \bf $G_{17}(2190)$****}} & \multicolumn{5}{c}{\raisebox{-2.0ex}{\large \bf $G_{19}(2250)$****}}\\[2.5ex]
		\cline{1-5} \cline{6-10}
		$2222(15)$        & $442(40)$         & $2162$            & $407$       & This~Work  & $2200(10)$        & $343(51)$         & $2127$            & $262$      & This~Work\\
		 --               &  --              & $2074$            & $327$       & R\"onchen 15B   &     --        &    --         & $2062$            & $403$      & R\"onchen 15B\\
		--               &   --             &   --              &   --        & Workman 12   & --     &     --         &     --        &  --      & Workman 12\\
		$2180(20)$       & $335(40)$        & $2150(25)$        & $330(30)$   & Anisovich 12   & $2280(40)$        &$520(50)$         & $2195(45)$        & $470(50)$ & Anisovich 12 \\
				 \hline	\hline
								
	\end{tabular}
			\caption{ Comparison of $D_{15}$, $F_{15}$, $F_{17}$, $G_{17}$, and $G_{19}$ resonance masses, widths, and pole positions for isospin-1/2 amplitudes. Star rating is that found in the RPP~\cite{pdg}. Comparisons are made with works by R\"onchen 15b~\cite{julich15}, Anisovich 12~\cite{anisovich12}, and SAID~\cite{workman12}.}
			\label{MassPole2}	
	\end{table*}
	
	\begin{table*}[hbtp]
		
		\begin{tabular}{@{\extracolsep{1pt}}cccclCcccl@{}}
			\hline \hline
			\raisebox{-0.5ex}{\specialcell {$A_{\frac{1}{2}}^p$ \\(${\rm GeV}^{-1/2}$)~~}}& \raisebox{-0.5ex}{\specialcell {$A_{\frac{1}{2}}^n$ \\(${\rm GeV}^{-1/2}$)~~}} & \raisebox{-0.5ex}{\specialcell {$A_{\frac{3}{2}}^p$ \\(${\rm GeV}^{-1/2}$)~~}} & \raisebox{-0.5ex}{\specialcell {$A_{\frac{3}{2}}^n$ \\(${\rm GeV}^{-1/2}$)~}} &
			\raisebox{-0.5ex}{\bf Analysis} &
			\raisebox{-0.5ex}{\specialcell {$A_{\frac{1}{2}}^p$ \\(${\rm GeV}^{-1/2}$)~~}}& \raisebox{-0.5ex}{\specialcell {$A_{\frac{1}{2}}^n$ \\(${\rm GeV}^{-1/2}$)~~}} & \raisebox{-0.5ex}{\specialcell {$A_{\frac{3}{2}}^p$ \\(${\rm GeV}^{-1/2}$)~~}} & \raisebox{-0.5ex}{\specialcell {$A_{\frac{3}{2}}^n$ \\(${\rm GeV}^{-1/2}$)~}} & \raisebox{-0.5ex}{\bf ~Analysis}\\[1.1ex] \hline			
			\multicolumn{5}{c}{\raisebox{-2.0ex}{\large \bf $S_{11}(1535)$****}} & \multicolumn{5}{c}{\raisebox{-2.0ex}{\large \bf $S_{11}(1650)$****}}\\[2.5ex]
			\cline{1-5} \cline{6-10}
			
			$+0.107(3)$       &        $-0.055(6)$   &              &           & This~Work  & $+0.048(3)$       &  $+0.001(6)$         &              &           & This~Work\\
			Not included     &                  &                   &            & R\"onchen 15B   & Not included     &                  &                   &            & R\"onchen 15B\\
			$+0.128(4)$      &                  &                   &            & Workman 12   & $+0.055(30)$     &                  &                   &            & Workman 12\\
			$+0.105(10)$   		& $-0.093(11)$  &                  &		    & Anisovich   &   $+0.033(7)$      & $+0.025(20)$    &                 &                   & Anisovich\\
			\hline 
			\multicolumn{5}{c}{\raisebox{-2.0ex}{\large \bf $S_{11}(1895)$****}} & \multicolumn{5}{c}{\raisebox{-2.0ex}{\large \bf $P_{11}(1440)$****}}\\[2.5ex]
			\cline{1-5} \cline{6-10}
			$+0.017(5)$      & $+0.002(13)$        &              &           & This~Work  & $-0.091(7)$              &        $+0.013(12)$        &              &           & This~Work\\
			--     &      --          &     --        &   --       & R\"onchen 15B   &     --        &    --           & --            & --      & R\"onchen 15B\\
			--     &      --          &     --        &   --       & Workman 12   & $-0.056(1)$       &                  &                   &            & Workman 12\\
			$-0.011(6)$      & $+0.013(6)$       &              &                 & Anisovich   & $-0.061(8)$      & $+0.043(12)$  &                       &             & Anisovich \\
			\hline 
			\multicolumn{5}{c}{\raisebox{-2.0ex}{\large \bf $P_{11}(1710)$****}} & \multicolumn{5}{c}{\raisebox{-2.0ex}{\large \bf $P_{11}(1880)$***}}\\[2.5ex]
			\cline{1-5} \cline{6-10}
			$+0.014(8)$      &   $+0.0053(3)$          &              &           & This~Work  & $+0.119(15)$      &        $+0.016(10)$      &              &           & This~Work\\
			--       &      --        & --    & --      & R\"onchen 15B    &         --         &      --            & --            & --      & R\"onchen 15B\\
			--     &       --           &        --           &     --       & Workman 12   & --     &        --          &     --         &   --         & Workman 12\\
			$+0.052(15)$      & $-0.40(20)$     &                  &           & Anisovich   & $-0.013(3)$         & $+0.034(11)$ &  &   & Anisovich 12a,b \\
			\hline 
			\multicolumn{5}{c}{\raisebox{-2.0ex}{\large \bf $P_{11}(2100)$***}} & \multicolumn{5}{c}{\raisebox{-2.0ex}{\large \bf $P_{13}(1720)$****}}\\[2.5ex]
			\cline{1-5} \cline{6-10}
			$+0.032(14)$     &    $+0.026(13)$  &              &           & This~Work  & $+0.068(4)$      &        $-0.064(6)$    &	$+0.028(3) $            &    $-0.004(6)$       & This~Work\\
			--	&      --       & --            & --    & R\"onchen 15B   &   --       &   --      & --            & --      & R\"onchen 15B\\
			--           & --         &      --          &   --         & Workman 12   & $+0.095(2)$       &                  &    $-0.048(2)$               &            & Workman 12\\
			--        & --       & --         &  --  & Anisovich 12   & $+0.110(45)$         & $-0.080(50)$  & $+0.150(30)$    & $-0.140(65)$  & Anisovich \\
			\hline 
			\multicolumn{5}{c}{\raisebox{-2.0ex}{\large \bf $P_{13}(1900)$****}} & \multicolumn{5}{c}{\raisebox{-2.0ex}{\large \bf $P_{13}(2040)$*}}\\[2.5ex]
			\cline{1-5} \cline{6-10} 
			$+0.040(4)$       &  $+0.007(14)$      & $-0.094(7)$             &  $+0.007(11)$         & This~Work  & $+0.038(7)$   &  $+0.025(21)$        & $+0.078(10)$          &  $-0.091(20)$         & This~Work\\
			--     &    --         &     --         &     --    & R\"onchen 15B  & --        & --        & --            & --      & R\"onchen 15B\\
			--     &    --         &     --         &     --    & Workman 12   & --        & --         & --            & --     & Workman 12 \\
			$+0.026(15)$     & $+0.000(30)$   &   $-0.065(30)$  & $-0.060(45)$  & Anisovich   & --        & --         & --            & --      & Anisovich \\ \hline	
			\multicolumn{5}{c}{\raisebox{-2.0ex}{\large \bf $D_{13}(1520)$****}} & \multicolumn{5}{c}{\raisebox{-2.0ex}{\large \bf $D_{13}(1700)$***}}\\[2.5ex]
			\cline{1-5} \cline{6-10}
			$-0.034(3)$    & $-0.072(3)$            & $+0.142(3)$  	     &   $-0.123(6)$        & This~Work  & $+0.032(5)$        &  $+0.005(11)$    & 	$+0.034(6)$   & $-0.094(17)$    & This~Work\\
			--             &   --             & --            & --       & R\"onchen 15B   & --     &    --            &    --             &     --     & R\"onchen 15B\\
			$-0.019(2)$     &                  &      $+0.141(2)$             &            & Workman 12   & --     &    --            &    --             &     --     & Workman 12\\
			$-0.022(4)$      &   $-0.049(8)$	  &   $+0.131(10)$    &  $-0.113(12)$		 & Anisovich   & $+0.041(17)$      &  $+0.025(10)$     &  $-0.034(13)$   &  $-0.032(18)$   & Anisovich \\ \hline
			\multicolumn{5}{c}{\raisebox{-2.0ex}{\large \bf $D_{13}(1875)$***}} & \multicolumn{5}{c}{\raisebox{-2.0ex}{\large \bf $D_{13}(2120)$***}}\\[2.5ex]
			\cline{1-5} \cline{6-10}
			$-0.013(8)$    &  $+0.050(9)$ & $-0.093(9)$  &   $+0.141(22)$    & This~Work  & $+0.047(9)$   &  $-0.020(13)$            &	$+0.001(7)$ &       $-0.00(2)$    & This~Work\\
			--     &    --            &    --             &     --     & R\"onchen 15B   & --     &    --            &    --             &     --     & R\"onchen 15B\\
			--     &    --            &    --             &     --     & Workman 12   & --     &    --            &    --             &     --     & Workman 12\\
			$+0.018(10)$     & $+0.010(6)$ &$-0.009(5)$     & $-0.020(15)$ & Anisovich   &               &	$+0.110(45)$       &                    & $+0.040(30)$    & Anisovich 13b \\ \hline		\hline									
		\end{tabular}
		
		\caption{\label{helicity12_SPD} Comparison of $S_{11}$, $P_{11}$, $P_{13}$, and $D_{13}$ helicity-1/2 and 3/2 amplitudes for both the proton and neutron. Star rating is that found in the RPP~\cite{PDG2018}. Comparisons are made with works by R\"onchen 15b~\cite{julich15}, Anisovich 12~\cite{anisovich12}, and SAID~\cite{workman12}.}
	\end{table*}

	\begin{table*}[hbtp]	
		\begin{tabular}{@{\extracolsep{1pt}}cccclCcccl@{}}
			\hline \hline
			\raisebox{-0.5ex}{\specialcell {$A_{\frac{1}{2}}^p$ \\(${\rm GeV}^{-1/2}$)~~}}& \raisebox{-0.5ex}{\specialcell {$A_{\frac{1}{2}}^n$ \\(${\rm GeV}^{-1/2}$)~~}} & \raisebox{-0.5ex}{\specialcell {$A_{\frac{3}{2}}^p$ \\(${\rm GeV}^{-1/2}$)~~}} & \raisebox{-0.5ex}{\specialcell {$A_{\frac{3}{2}}^n$ \\(${\rm GeV}^{-1/2}$)~}} &
			\raisebox{-0.5ex}{\bf Analysis} &
			\raisebox{-0.5ex}{\specialcell {$A_{\frac{1}{2}}^p$ \\(${\rm GeV}^{-1/2}$)~~}}& \raisebox{-0.5ex}{\specialcell {$A_{\frac{1}{2}}^n$ \\(${\rm GeV}^{-1/2}$)~~}} & \raisebox{-0.5ex}{\specialcell {$A_{\frac{3}{2}}^p$ \\(${\rm GeV}^{-1/2}$)~~}} & \raisebox{-0.5ex}{\specialcell {$A_{\frac{3}{2}}^n$ \\(${\rm GeV}^{-1/2}$)~}} & \raisebox{-0.5ex}{\bf ~Analysis}\\[1.1ex] \hline			
			\multicolumn{5}{c}{\raisebox{-2.0ex}{\large \bf $D_{15}(1675)$****}} & \multicolumn{5}{c}{\raisebox{-2.0ex}{\large \bf $D_{15}(2060)$***}}\\[2.5ex]
			\cline{1-5} \cline{6-10}
			
			$+0.026(2)$        &   $-0.069(5)$   &  	$+0.005(2)$         &   $-0.031(5)$        & This~Work  & $-0.019(5)$     &   $+0.069(17)$      &   $+0.039(5)$     &    $-0.023(20)$       & This~Work\\
			--              &   --             & --            & --        & R\"onchen 15B   & --     &     --         &     --        &  --      & R\"onchen 15B\\
			$+0.013(1)$     &     --        &    $+0.016(1)$               &     --     & Workman 12   & --     &     --         &     --        &  --      & Workman 12\\
			$+0.024(3)$         & $-0.060(7)$ & $+0.025(7)$    & $-0.088(10)$    & Anisovich   & $+0.067(15)$    & $+0.025(11)$      & $+0.055(20)$      &  $-0.037(17)$  & Anisovich \\
			\hline 
			\multicolumn{5}{c}{\raisebox{-2.0ex}{\large \bf $F_{15}(1680)$****}} & \multicolumn{5}{c}{\raisebox{-2.0ex}{\large \bf $F_{15}(1860)$**}}\\[2.5ex]
			\cline{1-5} \cline{6-10}
			$-0.026(4)$  &     $+0.005(4)$         &	$+0.112(5)$ &     $-0.061(4)$      & This~Work    & $-0.022(20)$         & $+0.021(29)$     & $-0.032(34)$      & $+0.070(35)$      & This~Work\\
			--            &   --             & --            & --      & R\"onchen 15B  &  --     &     --         &     --        &  --      & R\"onchen 15B\\
			$-0.007(2)$     &     --        &   $+0.140(2)$          &     --       & Workman 12   & --     &     --         &     --        &  --      & Workman 12\\
			$-0.013(3)$         & $+0.034(6)$ & $+0.135(6)$    & $+0.044(9)$            & Anisovich   & $-0.019(11)$         & $+0.021(13)$     & $+0.048(18)$      & $+0.034(17)$   & Anisovich \\
			\hline 
			\multicolumn{5}{c}{\raisebox{-2.0ex}{\large \bf $F_{17}(1990)$**}} & \multicolumn{5}{c}{\raisebox{-2.0ex}{\large \bf $F_{17}(2200)$ new}}\\[2.5ex]
			\cline{1-5} \cline{6-10}
			$+0.006(3)$      &    $-0.027(24)$   &	$-0.055(8)$     & $+0.051(20)$  & This Work & $-0.000(5)$      &   $+0.035(36)$       &   $-0.128(13) $        &  $+0.031(31)$        & This~Work\\
			--               &  --              & --          & --      & R\"onchen 15B   & --     &     --         &     --        &  --      & R\"onchen 15B\\
			--               & --               &   --              &   --       & Workman 12   & --     &     --         &     --        &  --      & Workman 12\\
			$+0.040(12)$     &  $-0.045(20)$   &   $+0.057(12)$    & $-0.052(27)$ & Anisovich   & --     &     --         &     --        &  --      & Anisovich 12 \\
			\hline 
			\multicolumn{5}{c}{\raisebox{-2.0ex}{\large \bf $G_{17}(2190)$****}} & \multicolumn{5}{c}{\raisebox{-2.0ex}{\large \bf $G_{19}(2250)$****}}\\[2.5ex]
			\cline{1-5} \cline{6-10}
			$+0.001(2)$      &  $-0.01(2)$     &	$+0.015(3)$      &  $-0.023(22)$     & This~Work  & $+0.0006(37)$              &         &     $+0.013(4)$              &            & This~Work\\
			--               &  --              & --            & --       & R\"onchen 15B   &     --        &    --         & --        &    --      & R\"onchen 15B\\
			--               &   --             &   --              &   --        & Workman 12   & --     &     --         &     --        &  --      & Workman 12\\
			$-0.065(8)$      & $-0.015(13)$    &   $+0.035(17)$      & $-0.052(27)$  & Anisovich   & --        &    --         & --        &    -- & Anisovich 12 \\
			\hline	\hline
			
		\end{tabular}

		\caption{\label{Helicity12_DFG} Comparison of $D_{15}$, $F_{15}$, $F_{17}$, $G_{17}$, and $G_{19}$ helicity-1/2 and 3/2 amplitudes for both the proton and neutron. Star rating is that found in the RPP~\cite{PDG2018}. Comparisons are made with works by R\"onchen 15b~\cite{julich15}, Anisovich 12~\cite{anisovich12}, and SAID~\cite{workman12}.}	
	\end{table*}

\subsection*{$S_{11}$}
This amplitude required four resonances within the fitting region. The first two are the well known $S_{11}(1535)$ and $S_{11}(1650)$ and are clearly seen in $\pi N$, $K\Lambda$, and $\eta N$ photoproduction. The properties of the third state, $S_{11}(1895)$, especially its mass, were primarily constrained by the $\pi N \rightarrow \pi N$ and the $\pi N \rightarrow \eta N$ reactions and it was the $\pi N \rightarrow \eta N$ reaction that required the resonance.  The  $S_{11}(1895)$ was listed as a 2-star resonance in the 2016 edition of the RPP \cite{pdg}, but it was promoted to a 4-star resonance in the 2018 edition \cite{PDG2018}.  A fourth resonance at 2400~MeV was used to constrain the high-energy behavior of the $K\Lambda$ channels and remains inconclusive. At this stage of the analysis, its parameters are not reliable and are not quoted. 

In general, our parameter values for the $S_{11}(1535)$ and $S_{11}(1650)$ are in agreement with results from other works. The primary exception is the $S_{11}(1535)$ helicity-1/2 coupling found in this work, which is larger than the result by Shrestha \textit{et al.} ~\cite{ShresthaED} but is now in agreement with other more recent results. However, results for the $S_{11}(1895)$ are still not in good agreement between the different groups. For instance, a few groups find a width less than 150~MeV, which is quite narrow, while this and other works find a width in excess of 400~MeV, which is quite large. While the helicity-1/2 couplings show different signs, early indications suggest the resonance has a weak photocoupling. Our fit of the $S_{11}$ amplitudes contained no dummy channels, meaning that $S_{11}$ inelasticity can be explained by the measured reactions.

\subsection*{$P_{11}$}
$P_{11}$ required four resonances, including the well-known Roper resonance $P_{11}(1440)$. The Roper resonance shows up in this analysis with a lower mass and width than most current groups seem to find, as well as a larger helicity-1/2 coupling about twice as large. The results for the $P_{11}(1710)$ are also quite different from other groups because it was never clearly seen in any reaction. In this work it shows up as a clear resonance bump in the reaction \GPEP \ with a well-determined mass and width. Its mass in this work is smaller than that found by other works, while its width is similar to more recent results. Only BnGa finds a large helicity coupling to the resonance (both to the proton and neutron). The agreement between groups for the $P_{11}(1880)$ resonance is also poor. The  $P_{11}(1880)$ was listed as a 2-star resonance in the 2016 edition of the RPP \cite{pdg}, but it was promoted to a 3-star resonance in the 2018 edition \cite{PDG2018}.  This work finds a strong helicity-1/2 coupling to the proton for the $P_{11}(1880)$, which disagrees with other results. The large resonance coupling was a stable feature of our analysis and was suggested in both the \GPEP \ and \GPKL \ reactions. Evidence for a strong coupling is strengthened by the fact that even when the amplitude was started small and then varied, plots of the modulus showed a distinct bump, which is a signature of a resonance. A fourth $P_{11}$ resonance at 2200~MeV was included to help explain the high-energy behavior, but nothing conclusive can be said about its properties.   This state is listed in the tables as $P_{11}$(2100).  The  $P_{11}(2100)$ was listed as a 1-star resonance in the 2016 edition of the RPP \cite{pdg}, but it was promoted to a 3-star resonance in the 2018 edition \cite{PDG2018}. Our fit of the $P_{11}$ amplitudes used two $\rho\Delta$ dummy channels.

\subsection*{$P_{13}$}
$P_{13}$ required three resonances. It is also the dominant amplitude above the $S_{11}(1650)$ resonance for the reactions \GPKL \ and \PINK. The mass and width of the $P_{13}(1720)$ were determined by both \GPEP \, and \GPKL. This is in stark contrast to other analyses that find little or no need for $P_{13}$ in the reactions involving $\eta N$. For the $P_{13}(1720)$, the helicity-$3/2$ coupling to the proton is still in poor agreement between different groups as some works find a small negative value while others (including this work) find a small positive value. Also, BnGa found a large negative helicity-3/2 coupling to the neutron, while other groups (including this work) find a small negative value. The $P_{13}(1900)$ was first seen in the $\pi \pi N$ channels~\cite{Manley95}, but its properties are constrained by \GPKL. The  $P_{13}(1900)$ was listed as a 3-star resonance in the 2016 edition of the RPP \cite{pdg}, but it was promoted to a 4-star resonance in the 2018 edition \cite{PDG2018}.  Its mass and helicity parameters are now in good agreement between groups, but its width shows disagreement between this work and others such as \cite{ShresthaED}. A third $P_{13}$ resonance at 2244~MeV was used to fit the data above 2000~MeV for the reaction \GPPN. The dummy channels for our fit of the $P_{13}$ amplitudes were $\rho\Delta$, $\omega N$, and $K \Sigma$.

\subsection*{$D_{13}$}
$D_{13}$ required four resonances. The $D_{13}$(1520) is clearly seen in the $\pi N$ elastic and photoproduction reactions. For this reason, groups generally agree on its parameters. The $D_{13}(1700)$ resonance was initially seen in  $\pi N \rightarrow \pi \pi N$, but this work also finds evidence in the reactions \GPEP \ and \GNEN. Due to its lack of a strong coupling to a single channel, the resonance has a poorly determined mass and width. The $D_{13}(1875)$ resonance is hinted at in $\eta$ photoproduction but with poorly determined properties due to lack of data near 1875 MeV. Its mass in this work is higher than that found in other works except H\"ohler~\cite{hohler79} and its width and helicity couplings are in poor agreement among most groups with a width ranging from 180 to 900~MeV. A fourth $D_{13}$ resonance at 2353~MeV, listed in the tables as $D_{13}$(2120), was included due to some indication of its existence in the reaction \GPKL.  The  $D_{13}(2120)$ was listed as a 2-star resonance in the 2016 edition of the RPP \cite{pdg}, but it was promoted to a 3-star resonance in the 2018 edition \cite{PDG2018}.  No dummy channels were used in our fit of the $D_{13}$ amplitudes.

\subsection*{$D_{15}$}
This partial wave required two resonances, the $D_{15}(1675)$ and the $D_{15}(2060)$. The $D_{15}(1675)$  has well-defined parameters due to the resonance having a strong coupling to both the $\pi N$ channel and $\pi \pi N$ channels. It also contains very little background contributions in most reactions. The main exceptions are the photoproduction reactions on the proton. This is due to the Moorhouse selection rule~~\cite{Moorhouse}, which states that the first $D_{15}$ resonance should not couple to $\gamma p$. The $D_{15}(2060)$ is seen in the data for the reaction \GPKL \ and was necessary to obtain a good fit to differential cross-section data above 2000~MeV. The  $D_{15}(2060)$ was listed as a 2-star resonance in the 2016 edition of the RPP \cite{pdg}, but it was promoted to a 3-star resonance in the 2018 edition \cite{PDG2018}.  The only dummy channel for our fit of the $D_{15}$ amplitudes was a $\rho\Delta$ channel.

\subsection*{$F_{15}$}
$F_{15}$ needed three resonances, including the $F_{15}(1680)$ and $F_{15}(1860)$. The $F_{15}(1680)$ is well determined from pion reactions and groups agree on its parameters. The $F_{15}(1860)$ is less clear but necessary to fit the high-energy behavior of $\eta$ photoproduction. There is also a hint of a resonance in $\pi N \rightarrow \pi N$  where a small bump does appear. However, a good fit of the bump proved difficult as improvements in the fit to the imaginary part degraded fits to the real part. This may be one reason groups tend to agree on its mass but not its width. A third resonance at 2320~MeV was clear in the magnetic amplitude for the reaction \GPKL. No dummy channels were used in our fit of the $F_{15}$ amplitudes.

\subsection*{$F_{17}$}
 $F_{17}$ needed two resonances, namely the $F_{17}(1990)$ and $F_{17}(2200)$. The $F_{17}(1990)$ has poorly determined parameters and was not conclusively seen in any reaction, although there are hints that it is necessary in \GPEP, \GPKL, and perhaps $\pi N \rightarrow \pi N$. The $F_{17}(2200)$ is a new state that was added to fit the indication of a higher-lying resonance in $\pi N \rightarrow \pi N$ where the imaginary part starts increasing above 2000~MeV. Based on the single-energy solution, it appears it will peak just above 2300~MeV. This work also finds the $F_{17}(2200)$ has a strong coupling to $K \Lambda$. This is in agreement with quark-model predictions from ~\cite{Capstick98}. A reliable determination of its parameters would most likely require data up to 2400~MeV in a number of reactions, including $\pi N \rightarrow \pi N$. This amplitude was also critical for describing the forward-angle shape of the differential cross section at energies above 1800~MeV for the reaction \PINE. The dummy channels in our fit of the $F_{17}$ amplitudes were $K \Sigma$, $\omega N$, and $\rho \Delta$.

\subsection*{$G_{17}$ and $G_{19}$}
The $G_{17}(2190)$ and $G_{19}(2250)$ resonances were used in the higher amplitudes and are seen primarily in $\pi N \rightarrow \pi N$. Both resonances had negligible helicity coupling and are not seen in any photoproduction reaction. Groups generally agree on the resonance parameters for $G_{17}(2190)$ because it clearly appears in $\pi N \rightarrow \pi N$; however, the properties of $G_{19}(2250)$  show significant disagreement between groups. The only agreement is that its mass is most likely above 2200~MeV. An $\omega N$ dummy channel was used in our fit of the $G_{17}$ amplitudes while a  $\rho N$ dummy channel was used in our fit of the $G_{19}$ amplitudes.



\subsection{\bf Results for Isospin-$3/2$ Amplitudes}
\label{Iso32Results}
The following section discusses results for the isospin-3/2 amplitudes. Table \ref{MassPole3} contains the masses ($M$), widths ($\Gamma$), and pole positions for each isospin-3/2 resonance with uncertainties on the last reported significant figure shown in parentheses. Table \ref{Iso32Helicity} shows helicity-3/2 couplings for each resonance. Table \ref{PartialWidths3} of the appendix shows the partial widths ($\Gamma_i$), branching fractions (${\cal B}_i$), and resonant amplitudes ($\sqrt{xx_i}$) for each amplitude's included channels.

\begin{table*}[hbtp]
	
	\begin{tabular}{@{\extracolsep{1pt}}cccclCcccl@{}}
		\hline \hline
		\raisebox{-0.5ex}{~~\specialcell {Mass \\(MeV)}~~}& \raisebox{-0.5ex}{~~\specialcell {Width \\(MeV)}~~} & \raisebox{-0.5ex}{~~\specialcell {Pole \\(Re)}~~} & \raisebox{-0.5ex}{~~\specialcell {Pole \\(-2Im)}~~} &
		\raisebox{-0.5ex}{\bf ~~Analysis} &
		\raisebox{-0.5ex}{~~~\specialcell {Mass \\(MeV)}~~~}& \raisebox{-0.5ex}{~~\specialcell {Width \\(MeV)}~~} & \raisebox{-0.5ex}{~~\specialcell {Pole \\(MeV)}~~} &
		\raisebox{-0.5ex}{~~\specialcell {Pole \\(MeV)}~~} &\raisebox{-0.5ex}{\bf ~~Analysis}\\[1.1ex] \hline			
		\multicolumn{5}{c}{\raisebox{-2.0ex}{\large \bf $S_{31}(1620)$****}} & \multicolumn{5}{c}{\raisebox{-2.0ex}{\large \bf $S_{31}(1900)$***}}\\[2.5ex]
		\cline{1-5} \cline{6-10}
		
		$1589(3)$        & $107(7)$         & $1577$            & $101$      & This~Work  & $1989(22)$       & $457(60)$          & $1957$            & $447$      & This~Work\\
		&                  & $1600$            & $65$       & R\"onchen 15B   & --        & --         & --            & --      & R\"onchen 15B\\
		$1615$           & $147$            &                   &            & Workman 12   & --        & --         & --            & --      & Workman 12 \\
		$1600(8)$        & $130(11)$        & $1597(4)$         & $130(9)$   & Anisovich 12   & $1840(30)$       & $300(45)$          & $1845(25)$        & $300(45)$  & Anisovich 12 \\
	
		\hline 
		\multicolumn{5}{c}{\raisebox{-2.0ex}{\large \bf $P_{31}(1910)$****}} & \multicolumn{5}{c}{\raisebox{-2.0ex}{\large \bf $P_{31}(2250)$ new}}\\[2.5ex]
		\cline{1-5} \cline{6-10}
		$1846(18)$       & $260(57)$         & $1801$            & $224$      & This~Work  & $2250(30)$        & $320(120)$         & $2250$            & $320$      & This~Work\\
		&                   & $1799$            & $648$      & R\"onchen 15B   & --        & --         & --            & --      & R\"onchen 15B\\
		--        & --                 & --                & --      & Workman 12   & --        & --         & --            & --      & Workman 12\\
		$1860(40)$       & $350(55)$         & $1850(40)$        & $350(45)$  & Anisovich 12   & --        & --         & --            & --      & Anisovich 12 \\
		\hline 
		\multicolumn{5}{c}{\raisebox{-2.0ex}{\large \bf $P_{33}(1232)$****}} & \multicolumn{5}{c}{\raisebox{-2.0ex}{\large \bf $P_{33}(1600)$****}}\\[2.5ex]
		\cline{1-5} \cline{6-10}
		$1230.8(4)$      & $110.9(8)$       & $1212.4$          & $96.8$     & This~Work  & $1664(16)$       & $322(46)$         & $1619$            & $295$      & This~Work\\
		--          &     --           & $1218$            & $92$       & R\"onchen 15B   &    --            &     --           & $1552$            & $350$      & R\"onchen 15B\\
		$1233$           & $119$            &      --           &  --        & Workman 12   & --        & --         & --            & --      & Workman 12\\
		$1228(2)$        & $110(3)$         & $1210.5(10)$      & $99(2)$    & Anisovich 12   & $1510(20)$       & $220(45)$        & $1498(25)$        & $230(50)$  & Anisovich 12 \\
		\hline 
		\multicolumn{5}{c}{\raisebox{-2.0ex}{\large \bf $P_{33}(1920)$***}} & \multicolumn{5}{c}{\raisebox{-2.0ex}{\large \bf $D_{33}(1700)$****}}\\[2.5ex]
		\cline{1-5} \cline{6-10} 
		$1976.0(49)$      & $509(170)$       & $1910$            & $472$      & This~Work  & $1720(5)$        & $226(14)$         & $1693$            & $213$      & This~Work\\
		--             &   --             & $1715$            & $882$      & R\"onchen 15B   &       --         &     --         & $1677$            & $305$      & R\"onchen 15B\\
		--               &      --         &      --            & --         & Workman 12   & $1695$           & $376$            &      --           &     --      & Workman 12\\
		$1900(30)$       & $310(60)$        & $1890(30)$        &  $300(60)$ & Anisovich 12   & $1715^{+30}_{-15}$ & $310^{+40}_{-15}$ & $1680(10)$     &  $305(15)$ & Anisovich 12 \\ \hline	
		\multicolumn{5}{c}{\raisebox{-2.0ex}{\large \bf $D_{33}(1940)$**}} & \multicolumn{5}{c}{\raisebox{-2.0ex}{\large \bf $D_{35}(1930)$***}}\\[2.5ex]
		\cline{1-5} \cline{6-10}
		$2137(13)$        & $400(43)$         & $2139$            & $400$      & This~Work  & $1988(32)$       & $500(160)$        & $1863$            & $260$      & This~Work\\
		--         &     --         &    --              &     --      & R\"onchen 15B   &         --    &       --        & $1836$            & $724$      & R\"onchen 15B\\
		--               &      --         &      --            & --         & Workman 12   & --               &      --         &      --            & --         & Workman 12\\
		$1995^{+105}_{-60}$ & $450(100)$    & $1990^{+100}_{-50}$  & $450(90)$ & Anisovich 12   & --               &      --         &      --            & --         & Anisovich 12 \\ \hline
		\multicolumn{5}{c}{\raisebox{-2.0ex}{\large \bf $F_{35}(1905)$****}} & \multicolumn{5}{c}{\raisebox{-2.0ex}{\large \bf $F_{37}(1950)$****}}\\[2.5ex]
		\cline{1-5} \cline{6-10}
		$1866(9)$        & $289(20)$         & $1819$            & $253$      & This~Work  & $1913(4)$        & $241(10)$     & $1871$            & $206$      & This~Work\\
		--             &   --             & $1795$            & $247$      & R\"onchen 15B   &   --         &       --        &   --            &   --      & R\"onchen 15B\\
		$1858$           & $321$            &        --         &   --       & Workman 12   & --               &      --         &      --            & --         & Workman 12\\
		$1861(6)$        & $335(18)$        & $1805(10)$        & $300(15)$  & Anisovich 12   & $1915(6)$        & $246(10)$    & $1890(4)$         & $243(8)$   & Anisovich 12 \\ \hline \hline											
	\end{tabular}
	
	\caption{Comparison of resonance masses, widths, and pole positions for isospin-3/2 amplitudes.}
	\label{MassPole3}
\end{table*}

\subsection*{$S_{31}$} 
For this partial wave, two resonances were used. Our results for the $S_{31}(1620)$ are in good agreement with those of other groups despite the large repulsive background that appears at low energies in the $\pi N \rightarrow \pi N$ amplitude, which could potentially distort its properties.  The  $S_{31}(1900)$ was listed as a 2-star resonance in the 2016 edition of the RPP \cite{pdg}, but it was promoted to a 3-star resonance in the 2018 edition \cite{PDG2018}.  The $S_{31}(1900)$ mass and width found in this work are significantly larger than values found by other groups. The helicity couplings found in this work for both resonances now agree with other recent results except Shrestha \textit{et al.}~\cite{ShresthaED}. One surprise in the results from this work is the strength of the $S_{31}(1900)$ helicity-1/2 coupling. While the size of the coupling is large, there is no significant indication in the single-energy solution for pion photoproduction that it should be significantly smaller and an overall coupling was important to fit the differential cross-section data in the reaction \GPPN, which other groups are unable to fit~\cite{Hunt2017}.
No dummy channels were needed to fit the $S_{31}$ amplitudes.

\subsection*{$P_{31}$}
$P_{31}$ needed two resonances, the $P_{31}(1910)$ and a new high-mass state. This partial wave shows significant repulsive background in the $\pi N \rightarrow \pi N$ amplitude. The mass of the $P_{31}(1910)$ resonance was lower than that found by other recent analyses but in agreement with results by older analyses. One concern with this amplitude is the size of the helicity-1/2 coupling. The single-energy solution suggests that perhaps the overall coupling is too large, but the existence of a few points above the energy-dependent fit that also have smaller error bars makes it difficult to obtain any definitive conclusion. The $\pi N$ coupling to the resonance is in very good agreement with results by other groups~\cite{pdg}, which implies that there is no obvious reason to increase its value while decreasing the helicity coupling. A new resonance, $P_{31}(2250)$, was used to fit the $\pi N \rightarrow \pi N$ amplitude at energies above 2000~MeV and was also used to fit the real part of the pion photoproduction amplitude. A $\rho\Delta$ dummy channel was used for our fit of the $P_{31}$ amplitudes.

\subsection*{$P_{33}$}
$P_{33}$ needed three resonances, including the $P_{33}(1232)$ and the $P_{33}(1600)$. Our results for the $P_{33}(1232)$ are in good agreement with other groups, which is to be expected due to its dominance in the elastic and pion photoproduction reactions. The  $P_{31}(1600)$ was listed as a 3-star resonance in the 2016 edition of the RPP \cite{pdg}, but it was promoted to a 4-star resonance in the 2018 edition \cite{PDG2018}.  The $P_{33}(1600)$ was needed for the $\pi \pi N$ reactions and various groups disagree about its properties. A few groups such as BnGa and H\"ohler~\cite{hohler79} find masses near 1510~MeV, while other works, including this one, find a mass above 1600~MeV. The positive helicity couplings found in this work agree with results by Shrestha~\cite{ShresthaED} in sign and magnitude, while other groups find negative values.  A third resonance at 2250~MeV has parameter values that differ significantly between groups, which shows that its properties are still poorly determined. Figure \ref{Argand22} in the appendix showing the $\pi N$ elastic channel shows the reaction saturates the unitary bound nearly up to 1500~MeV where $\pi \pi N$ channels become important. We included $\rho \Delta$ and $K \Sigma$ as dummy channels for our fit of the $P_{33}$ amplitudes.

\subsection*{$D_{33}$}
$D_{33}$ needed two resonances. The $D_{33}(1700)$ is well known and our values for its mass and width agree well with prior analyses. In addition, our value for its helicity-1/2 coupling is in agreement with more recent results. This work found a second $D_{33}$ resonance at 2137~MeV. Its parameters in general differ from those of other works, and some groups, including SAID~\cite{Strakovsky15}, do not include a second resonance in their fits, despite this work having found significant evidence for its existence in \GPPN. Interestingly, the helicity-1/2 coupling found in this work agrees with the work by Sokhoyan~\cite{sokhoyan15}, but the helicity-3/2 coupling differs in sign. No dummy channels were needed for our fit of the $D_{33}$ amplitudes.

\subsection*{$D_{35}$}
This partial wave needed only the $D_{35}(1930)$ resonance. Its mass is similar to that found by other works except Arndt~\cite{arndt06}, while its width varies significantly among the different analyses. The helicity couplings also show differing signs and strengths among the different analyses. This work found a significant negative helicity-1/2 coupling to the resonance, while other groups have found a small coupling. A $\rho \Delta$ dummy channel was used for our fit of the $D_{35}$ amplitudes.

\subsection*{$F_{35}$} This partial wave needed the $F_{35}(1905)$ resonance and a higher-mass state.  The mass, width, and helicity couplings of $F_{35}(1905)$ are in good agreement among the different analyses, in part, because there is a clear indication for its existence in $\pi N \rightarrow \pi N$.  A second  $F_{35}$ state was needed at 2340~MeV to fit the high-energy behavior of the $\pi N \rightarrow \pi N$ amplitude and the suggestion of a structure appearing in pion photoproduction.  No dummy channels were needed in our fit of the $F_{35}$ amplitudes.

\subsection*{$F_{37}$}
$F_{37}$ needed two resonances, the $F_{37}(1950)$ and $F_{37}(2390)$. The $F_{37}(1950)$ has mass, width, and helicity couplings that are in good agreement among the different analyses and clearly appears in the $\pi N \rightarrow \pi N$ amplitude. The second resonance is located at 2390~MeV and was used to constrain the amplitudes at high energies, but there currently is only weak evidence for its existence. This state is listed as a 1-star resonance in the 2018 edition of the RPP \cite{PDG2018}.  We included $\rho \Delta$ and $K \Sigma$ as dummy channels in our fit of the $F_{37}$ amplitudes.

\subsection*{$G_{37}$ and $G_{39}$}
Our fits of the $G_{37}$ and $G_{39}$ amplitudes included only a single resonance with masses of 2330~MeV and 2300~MeV, respectively. 
Due to their high masses, their individual parameters are poorly determined and are not quoted.

\begin{table*}[hbtp]
	
	\begin{tabular}{@{\extracolsep{20pt}}cclccl@{}}
		\hline \hline
		\raisebox{-0.5ex}{\specialcell {$A_{\frac{1}{2}}^N$ \\(${\rm GeV}^{-1/2}$)~~}}  & \raisebox{-0.5ex}{\specialcell {$A_{\frac{3}{2}}^N$ \\(${\rm GeV}^{-1/2}$)~~}} & 
		\raisebox{-0.5ex}{\bf Analysis} &
		\raisebox{-0.5ex}{\specialcell {$A_{\frac{1}{2}}^N$ \\(${\rm GeV}^{-1/2}$)~~}}& \raisebox{-0.5ex}{\specialcell {$A_{\frac{3}{2}}^N$ \\(${\rm GeV}^{-1/2}$)~~}}  & \raisebox{-0.5ex}{\bf ~Analysis}\\[1.1ex] \hline	
		\multicolumn{3}{c}{\raisebox{-2.0ex}{\large \bf $S_{31}(1620)$****}} & \multicolumn{3}{c}{\raisebox{-2.0ex}{\large \bf $S_{31}(1900)$***}}\\[2.5ex]
		\cline{1-3} \cline{4-6}
		
		$+0.0124(7)$          &    --       & This~Work  & $+0.212(29)$    &    --     & This~Work\\
		        --      & --      & R\"onchen 15B   & --   & --      & R\"onchen 15B\\
		$+0.029(3)$    &   --         & Workman 12   & --      & --      & Workman 12 \\
		$+0.052(5)$   & 	--	 & Anisovich 12   & $+0.059(16)$  	 &         & Anisovich 12 \\
		\hline 
		\multicolumn{3}{c}{\raisebox{-2.0ex}{\large \bf $P_{31}(1910)$****}} & \multicolumn{3}{c}{\raisebox{-2.0ex}{\large \bf $P_{31}(2250)$ new}}\\[2.5ex]
		\cline{1-3} \cline{4-6}
		$+0.203(56)$      &     --    & This~Work  & $-0.054(28)$      &     --      & This~Work\\
		--         & --           & R\"onchen 15B   & --    & --      & R\"onchen 15B\\
		--        & --      & Workman 12   & --     & --      & Workman 12\\
		$+0.022(9)$    &         & Anisovich 12   & --            & --      & Anisovich 12 \\
		\hline 
		\multicolumn{3}{c}{\raisebox{-2.0ex}{\large \bf $P_{33}(1232)$****}} & \multicolumn{3}{c}{\raisebox{-2.0ex}{\large \bf $P_{33}(1600)$****}}\\[2.5ex]
		\cline{1-3} \cline{4-6}
		$-0.146(2)$    &  $-0.250(2)$  & This~Work  & $+0.0082(14)$ &   $+0.048(14)$    & This~Work\\
		--         &    --       & R\"onchen 15B   &  --           & --      & R\"onchen 15B\\
		$-0.139(2)$  &  $-0.262(3)$  & Workman 12   & --            & --      & Workman 12\\
		$-0.131(4)$   &  $-0.254(5)$ &   Anisovich 12   & $-0.050(9)$    & $-0.040(12)$     &  Anisovich 12 \\
		\hline 
		\multicolumn{3}{c}{\raisebox{-2.0ex}{\large \bf $P_{33}(1920)$***}} & \multicolumn{3}{c}{\raisebox{-2.0ex}{\large \bf $D_{33}(1700)$****}}\\[2.5ex]
		\cline{1-3} \cline{4-6} 
		$-0.028(10)$     &      $-0.043(14)$    & This~Work  & $+0.156(17)$   & $+0.0125(16)$  & This~Work\\
		--             &   --      & R\"onchen 15B   &     --            & --      & R\"onchen 15B\\
		--         & --         & Workman 12   & $+0.105(5)$    &  $+0.092(4)$   & Workman 12\\
		$+0.130^{+30}_{-60}$   & $-0.115^{+25}_{-50}$ &  Anisovich 12   & $+0.160(20)$  & $+0.165(25)$       &  Anisovich 12 \\ \hline	
		\multicolumn{3}{c}{\raisebox{-2.0ex}{\large \bf $D_{33}(1940)$**}} & \multicolumn{3}{c}{\raisebox{-2.0ex}{\large \bf $D_{35}(1930)$***}}\\[2.5ex]
		\cline{1-3} \cline{4-6}
		$+0.1614(31)$       &    $-0.209(23)$  & This~Work  & $-0.043(8)$      & $-0.020(17)$  & This~Work\\
		--                     &    --      & R\"onchen 15B   &     --     & --      & R\"onchen 15B\\
		--                    & --         & Workman 12   & --              & --         & Workman 12\\
		--                     & --    & Anisovich 12   & --             & --         & Anisovich 12 \\ \hline
		\multicolumn{3}{c}{\raisebox{-2.0ex}{\large \bf $F_{35}(1905)$****}} & \multicolumn{3}{c}{\raisebox{-2.0ex}{\large \bf $F_{37}(1950)$****}}\\[2.5ex]
		\cline{1-3} \cline{4-6}
		$+0.077(10)$         & $-0.053(29)$             & This~Work  & $-0.047(2)$    &          $-0.074(2)$         & This~Work\\
		--                       & --      & R\"onchen 15B   &   --         &       --           & R\"onchen 15B\\
		$+0.019(2)$           &      $-0.038(4)$        & Workman 12   & $-0.083(4)$               &  $-0.096(4)$         & Workman 12\\
		$+0.025(5)$          & $-0.049(4)$          &  Anisovich 12   & $-0.071(4)$             & $-0.094(5)$           &  Anisovich 12 \\ \hline \hline											
	\end{tabular}
	\caption{ Comparison of $S_{31}$, $P_{31}$, $P_{33}$, $D_{33}$, $D_{35}$, $F_{35}$, and $F_{37}$ helicity-1/2 and 3/2 amplitudes for both the proton and neutron. Star rating is that found in the RPP~\cite{pdg}. Comparisons are made with works by R\"onchen 15b~\cite{julich15}, Anisovich 12~\cite{anisovich12}, and SAID~\cite{workman12}}
	\label{Iso32Helicity}
\end{table*}

\section{Summary and Conclusions}
An updated multichannel, partial-wave analysis was performed by including newly determined single-energy amplitudes for the reactions \GPEP, \GPKL, and $\gamma n \rightarrow \eta n$ in our energy-dependent fits of the various partial waves. The proton helicity coupling to the $S_{11}(1535)$  is now in agreement with results from other works. Also, a new $F_{17}$ resonance near 2200~MeV was needed to fit the \PIEL, \GPPN, and $\gamma p \rightarrow K^+ \Lambda$ reactions. This is consistent with quark-model predictions from \cite{capstick-roberts} that an $F_{17}$ resonance couples to $K \Lambda$. Additional data at energies above 2200~MeV are needed to both confirm its existence and determine its properties. In addition to our updated determination of resonance parameters, our fits yield a new energy-dependent solution for all the various partial-wave and multipole amplitudes. This energy-dependent solution provides an excellent description \cite{OURWORK1, OURWORK2} of the observables data used to determine the final single-energy amplitudes.


\begin{acknowledgments}
	The authors would like to thank Dr.\ Igor Strakovsky for supplying much of the photoproduction database used for our single-energy analyses. This work was supported in part by the U.S. Department of Energy, Office of Science, Office of Nuclear Physics Research Division, under Awards No.\ DE-FG02-01ER41194 and DE-SC0014323, and by the Department of Physics at Kent State University. 
	
\end{acknowledgments}

%
%
%
%
%
%
%
%
%
%
%
%
%
%
%
%
%
%
%
%
%
%
%
%
%
%
%
%
%
%

	
		
\clearpage
\appendix*
\section{Resonance Parameters and Argand Diagrams}
Tables \ref{PartialWidths1}, \ref{PartialWidths2}, and \ref{PartialWidths3} of the appendix list the partial widths ($\Gamma_i$), branching fractions (${\cal B}_i$), and resonant amplitudes ($\sqrt{xx_i}$) for the isospin-1/2 and isospin-3/2 amplitudes. Figures~\ref{Argand01} - \ref{Argand26} show Argand diagrams of the dimensionless energy-dependent amplitudes (solid black curves) fitted to the final single-energy results (data points). Small filled black circles mark the c.m.\ energies in which resonances were found.  The diagrams show the real and imaginary parts of the amplitudes as well as a polar plot of the amplitude from threshold up to 2100~MeV or 2300~MeV. The bottom right corner shows the reaction, the name of the amplitude, and for the photoproduction amplitudes whether it is an electric ($E$) or magnetic ($M$) multiple.  Note that for $I=1/2$ amplitudes, $S_{11}(E) = E_{0+}$, $P_{11}(M) = M_{1-}$, $P_{13}(E) = E_{1+}$, $P_{13}(M) = M_{1+}$, 
$D_{13}(E) = E_{2-}$, $D_{13}(M) = M_{2-}$, $D_{15}(E) = E_{2+}$, $D_{15}(M) = M_{2+}$,
$F_{15}(E) = E_{3-}$, $F_{15}(M) = M_{3-}$, $G_{17}(E) = E_{3+}$, $G_{17}(M) = M_{3+}$,
and similarly for $I=3/2$ amplitudes. For small amplitudes, the amplitude is shown after scaling. The scaling factor is shown after the amplitude name. Dummy channels for reactions without data or single-energy fits were included to satisfy $S$-matrix unitarity.  Numerical data for the dimensionless single-energy $\gamma p \rightarrow \eta p$, $\gamma n \rightarrow \eta n$, and $\gamma p \rightarrow K^+ \Lambda$ amplitudes, and for the updated $\pi^- p \rightarrow \eta n$ and $\pi^- p \rightarrow K^0 \Lambda$ amplitudes are available in the form of a data file~\cite{PAPS}.

\begin{table*}
\begin{tabular}{@{\extracolsep{15pt}}cccc@{\extracolsep{55pt}}c @{\extracolsep{15pt}}ccc@{}}
\hline \hline           
  \raisebox{-0.5ex}{\bf	Channel}  & \raisebox{-0.5ex}{\bf $\Gamma_i$ (MeV)} & \raisebox{-0.5ex}{ ${\cal B}_i$ }   & \raisebox{-0.5ex}{$\mathbf{\sqrt{xx_i}}$}   &  \raisebox{-0.5ex}{\bf	Channel}  & \raisebox{-0.5ex}{\bf $\Gamma_i$ (MeV)} & \raisebox{-0.5ex}{ ${\cal B}_i$ }   & \raisebox{-0.5ex}{$\mathbf{\sqrt{xx_i}}$}   \\ [1.1ex]
  \hline 
  		\multicolumn{4}{c}{\raisebox{-2.0ex}{\large \bf $S_{11}(1535)$****}} & \multicolumn{4}{c}{\raisebox{-2.0ex}{\large \bf $S_{11}(1650)$****}}\\[2.5ex]
  				\cline{1-4} \cline{5-8}
  
$\pi N$         & $62(3)$           & $42(2)$        & $+0.42(2)$    &$\pi N$         & $86(6)$            & $64(4)$          &  $+0.64(4)$\\
$\eta N$        &  $63(5)$          & $43(3)$        & $+0.43(1)$  &$\eta N$        &  $1.0(8)$          & $0.8(6)$          &  $+0.07(3)$\\
$K \Lambda$     &                   &                  &        &$K \Lambda$     &  $5(3)$          & $3.5(2)$          &  $-0.15(4)$\\    
$(\pi \Delta)_D$& $<1.7$          & $<1.1$         & $-0.043(35)$  &$(\pi \Delta)_D$& $<0.3$          & $<0.2$         &  $-0.01(8)$\\
$(\rho_3 N)_D$  &  $<0.5$         & $<0.3$         & $+.025(15)$  &$(\rho_3 N)_D$  &  $20(5)$           & $15(3)$           & $+0.31(3)$\\

$\rho_1 N$      &  $20(3)$          & $14(2)$          & $-0.24(2)$    &$\rho_1 N$      &  $<5$           & $1.8(1.7)$         & $+0.11(5)$\\
$\epsilon N$      & $<1.5$          & $<1$         & $-0.04(4)$   &$\epsilon N$    & $16(5)$             & $12(4)$          & $+0.28(4)$\\
$\pi N^*$       & $<0.01$           & $<0.01$            & $+0.003(2)$    &$\pi N^*$       & $3(2)$               & $2(1)$             & $+0.12(3)$\\
\hline
\multicolumn{4}{c}{\raisebox{-2.0ex}{\large \bf $S_{11}(1895)$****}} & \multicolumn{4}{c}{\raisebox{-2.0ex}{\large \bf $P_{11}(1440)$****}}\\[2.5ex]
  				\cline{1-4} \cline{5-8}
$\pi N$         & $39(18)$           & $8(4)$          &  $+0.08(4)$  &$\pi N$         & $153(10)$           & $59(2)$          &  $+0.59(2)$\\
$\eta N$        &  $174(52)$         & $37(9)$         &  $-0.18(5)$  &$\eta N$        &                    &                  &           \\
$K \Lambda$     &  $31(21)$          & $7(4)$          &  $+0.07(2)$  &$K \Lambda$     &                   &                   &           \\
$(\pi \Delta)_D$& $<49$           & $<10$          &  $+0.05(5)$  &$(\pi \Delta)_P$& $56(9)$            & $22(4)$         &  $+0.36(3)$\\
$(\rho_3 N)_D$  &  $105(45)$         & $23(9)$         & $+0.14(4)$  &$\rho_1 N$      &  $<0.003$          & $0.00(0)$          & $-0.00(2)$\\
$\rho_1 N$      &  $<85$          & $<18$          & $+0.08(5)$  &$\epsilon N$      & $41(9)$            & $16(3)$         & $+0.31(3)$\\
$\epsilon N$      & $<59$         & $<13$          & $-0.08(4)$  &     &          &              &  \\
$\pi N^*$       & $34(24)$           & $7(5)$             &  $-0.08(4)$  &  &          &             & \\ \hline
		\multicolumn{4}{c}{\raisebox{-2.0ex}{\large \bf $P_{11}(1710)$****}} & \multicolumn{4}{c}{\raisebox{-2.0ex}{\large \bf $P_{11}(1880)$***}}\\[2.5ex]
		\cline{1-4} \cline{5-8}			
$\pi N$         & $23(13)$           & $12(6)$          &  $+0.12(6)$    &$\pi N$         & $125(42)$           & $25(6)$          &  $+0.25(6)$\\
$\eta N$        &  $33(19)$             &  $17(8)$              & $-0.14(4)$  &$\eta N$        & $11(6)$            &   $2(1)$          &   $-0.07(2)$     \\      
$K \Lambda$     &  $3.5(3)$              &  $1.8(1.5)$             & $+0.05(2)$  &$K \Lambda$     & $11(5)$              & $2(1)$            & $-0.075(20)$    \\      
$(\pi \Delta)_D$& $55(21)$            & $28(9)$         &  $+0.19(4)$   &$(\pi \Delta)_D$& $57(31)$            & $11(6)$         &  $-0.17(5)$\\
$\rho_1 N$      &  $34(17)$          & $17(9)$          & $-0.14(5)$   &$\rho_1 N$      &  $160(62)$          & $32(13)$          & $+0.29(4)$\\
$\epsilon N$      & $<33$            & $<16$         & $-0.10(5)$   &$\epsilon N$      & $<45$             & $<9$            & $-0.09(7)$\\
						
	\hline		
		\multicolumn{4}{c}{\raisebox{-2.0ex}{\large \bf $P_{11}(2100)$***}} & \multicolumn{4}{c}{\raisebox{-2.0ex}{\large \bf $P_{13}(1720)$****}}\\[2.5ex]
		\cline{1-4} \cline{5-8}			
$\pi N$         & $117(58)$           & $21(11)$          &  $+0.21(11)$  &$\pi N$         & $41(4)$           & $18(2)$          &  $+0.178(16)$\\
$\eta N$        &  $<25$             &    $<4.7$           & $-0.06(5)$  &$\eta N$        & $8.7(1.6)$           &  $3.8(5)$             & $+0.082(7)$   \\
$K \Lambda$     & $<5.4$             &   $<1.0$           &  $+0.024(3)$ &$K \Lambda$     & $37(7)$          &    $16(3)$            &  $-0.17(1)$   \\
$(\pi \Delta)_D$& $<40$            & $<7.5$         &  $-0.06(11)$ &      &             &                &    \\   
$\rho_1 N$      &  $284(140)$          & $52(19)$          & $-0.33(8)$ &     &            &           & \\ 
$\epsilon N$      & $<190$            & $<35$         & $-0.17(12)$    	&  &               &              & \\
\hline							
			\multicolumn{4}{c}{\raisebox{-2.0ex}{\large \bf $P_{13}(1900)$****}} & \multicolumn{4}{c}{\raisebox{-2.0ex}{\large \bf $P_{13}(2040)$*}}\\[2.5ex]
			\cline{1-4} \cline{5-8} 
$\pi N$         & $5.7(2.9)$           & $1.9(1)$          &  $+0.019(10)$	&$\pi N$         & $89(25)$           & $16.7(1)$       &  $+0.17(4)$\\
$\eta N$        & $3.8(1.4)$             &  $1.3(5)$              & $-0.016(3)$ 		&$\eta N$        & $73(27)$              & $14$            & $-0.15(4)$ \\
$K \Lambda$     & $40(8)$            &    $13.7(3)$              & $-0.052(16)$ 	&$K \Lambda$     & $<0.7$             &  $<.04$        & $+0.004(29)$ \\
$\rho_1 N$      &  $94(20)$          & $32(7)$          & $+0.079(19)$	&$\rho_1 N$      &  $52(40)$          & $10(1)$         & $+0.127(4)$\\
		
	\hline \hline
				
\end{tabular}

\caption{\label{PartialWidths1}Below each resonance name are listed coupling constants ($\Gamma_i$), branching fractions (${\cal B}_i$), and resonant amplitudes ($\mathbf{\sqrt{xx_i}}$) for the channels listed in columns one and five. Star rating is that found in the \textit{Review of Particle Physics} (RPP)~\cite{PDG2018}. Table contains couplings to $S_{11}$, $P_{11}$, $P_{13}$ resonances included in the fits.}
\end{table*}

\begin{table*}
	\begin{tabular}{@{\extracolsep{15pt}}cccc@{\extracolsep{55pt}}c @{\extracolsep{15pt}}ccc@{}}
		\hline \hline           
		\raisebox{-0.5ex}{\bf	Channel}  & \raisebox{-0.5ex}{\bf $\Gamma_i$ (MeV)} & \raisebox{-0.5ex}{ ${\cal B}_i$ }   & \raisebox{-0.5ex}{$\mathbf{\sqrt{xx_i}}$}   &  \raisebox{-0.5ex}{\bf	Channel}  & \raisebox{-0.5ex}{\bf $\Gamma_i$ (MeV)} & \raisebox{-0.5ex}{ ${\cal B}_i$ }   & \raisebox{-0.5ex}{$\mathbf{\sqrt{xx_i}}$}   \\ [1.1ex]
		\hline 
		
\multicolumn{4}{c}{\raisebox{-1.8ex}{\large \bf $D_{13}(1520)$****}} & \multicolumn{4}{c}{\raisebox{-1.8ex}{\large \bf $D_{13}(1700)$***}}\\[2.5ex]
\cline{1-4} \cline{5-8}
$\pi N$         & $71(2)$           & $58.3(1.5)$          &  $+0.58(2)$		&$\pi N$         & $3.0(1)$           & $3.7(1)$          &  $+0.037(10)$\\
$\eta N$        & $0.041(8)$           & $0.03(1)$           & $+0.014(2)$ 	&$\eta N$        & $0.9(5)$           &   $1.1(6)$       & $+0.020(6)$ \\
$K \Lambda$     &                   &                    &   	&$K \Lambda$     & $1.1(5)$            &  $1.3(7)$       & $-0.022(6)$ \\
$(\pi \Delta)_S$ & $25(3)$            & $21(2)$         & $-0.35(2)$	&$(\pi \Delta)_S$ & $9(6)$            & $11(8)$         & $+0.06(2)$\\
$(\pi \Delta)_D$ & $7.2(1.2)$            & $6(1)$         & $-0.19(1)$	&$(\pi \Delta)_D$ & $10.4(6.5)$       & $13(5)$         & $+0.07(2)$\\
$(\rho_{3} N)_S$  & $17.1(1.9)$          & $14.1(1.5)$      & $-0.29(2)$	&$(\rho_{3} N)_S$ & $6(3)$            & $7.5(3.6)$       & $-0.05(2)$\\
$\epsilon N$      & $<0.9$            & $<0.7$          & $-0.04(3)$	&$\epsilon N$    & $50(10)$            & $62(9)$         & $+0.15(2)$\\		

\hline
\multicolumn{4}{c}{\raisebox{-1.8ex}{\large \bf $D_{13}(1875)$***}} & \multicolumn{4}{c}{\raisebox{-1.8ex}{\large \bf $D_{13}(2120)$***}}\\[2.5ex]
		\cline{1-4} \cline{5-8}	
		
$\pi N$         & $24(5)$             & $7.5(1)$         & $+0.075(14)$	&$\pi N$         & $97(14)$           & $19(2)$          &  $+0.19(2)$\\
$\eta N$        & $10.6(2.6)$         & $3.3(8)$         & $+0.050(8)$ 	&$\eta N$        & $16(12)$              & $3.1(2.4)$      & $-0.08(3)$ \\
$K \Lambda$     & $3.6(1.4)$          &  $1.1(4)$        & $+0.029(5)$ 	&$K \Lambda$     & $43(14)$              & $8.5(2.5)$      &  $-0.13(2)$ \\
$(\pi \Delta)_S$  & $<6$              & $<2$             & $+0.017(34)$	&$(\pi \Delta)_S$  & $125(59)$       & $25(11)$         & $-0.22(4)$\\
$(\pi \Delta)_D$  & $54(21)$          & $17(6)$          & $-0.11(2)$	&$(\pi \Delta)_D$ & $171(62)$        & $34(11)$         & $+0.26(5)$\\
$(\rho_{3} N)_S$  & $147(36)$         & $46(10)$         & $+0.19(2)$	&$(\rho_{3} N)_S$ & $<16$            & $<3$          & $+0.044(48)$\\
$\epsilon N$    & $78(27)$            & $24.3(8.6)$      & $-0.135(30)$		&$\epsilon N$      & $46(26)$       & $9(5)$         & $-0.13(4)$\\		
		\hline

		\multicolumn{4}{c}{\raisebox{-2.0ex}{\large \bf $D_{15}(1675)$****}} & \multicolumn{4}{c}{\raisebox{-2.0ex}{\large \bf $D_{15}(2060)$***}}\\[2.5ex]
		\cline{1-4} \cline{5-8}
$\pi N$         & $53(3)$           & $33(1)$          &  $+0.33(1)$	&$\pi N$         & $26(6)$           & $5.3(1.4)$        & $+0.05(1)$\\
$\eta N$        & $3.3(5)$           &  $2.0(3)$          &  $-0.082(7)$ 	&$\pi N$         & $26(6)$           & $5.3(1.4)$        & $+0.05(1)$\\
$K \Lambda$     & $<0.06$        &  $<0.04$           & $-0.007(5)$ 	&$K \Lambda$     & $76(29)$              & $15(5)$          & $+0.09(1)$ \\
$(\pi \Delta)_D$  & $94(6)$         & $58.3(2)$         & $+0.437(5)$	&$(\pi \Delta)_D$  & $74(30)$         & $15(6)$         & $+0.09(2)$\\
$\rho_{1} N$  &  $<0.3$          & $<0.2$          & $-0.017(11)$	&$\rho_{1} N$      &  $21(31)$        & $<10$          & $+0.047(36)$\\
$(\rho_{3} N)_D$  &  $0.6(4)$          & $0.4(3)$          & $-0.036(13)$	&$(\rho_{3} N)_D$    &  $70(43)$         & $14(9)$         & $-0.09(3)$\\

\hline 
\multicolumn{4}{c}{\raisebox{-1.8ex}{\large \bf $F_{15}(1680)$****}} & \multicolumn{4}{c}{\raisebox{-1.8ex}{\large \bf $F_{15}(1860)$**}}\\[2.5ex]
		\cline{1-4} \cline{5-8}

$\pi N$         & $84(2)$           & $68.0(1)$       & $+0.680(9)$	&  $\pi N$         & $30(5)$           & $8.0(1)$          & $+0.08(1)$\\
$\eta N$        & $0.11(3)$            & $0.09(2)$          & $+0.025(3)$ 	&$\eta N$        & $0.4(3)$             & $0.11(9)$        & $+0.009(4)$ \\
$K \Lambda$     & $0.00(0)$            & $0.00(0)$           & $-0.0008(12)$ 	&$K \Lambda$     & $ <0.03$          & $0.00(1)$           & $-0.0015(15)$ \\
$(\pi \Delta)_{P}$  & $16(2)$         & $13(1)$         & $-0.300(15)$  	&$(\pi \Delta)_{P}$  & $39(24)$         & $10(6)$         & $+0.09(3)$\\
$(\pi \Delta)_{F}$  & $<0.4$        & $<0.3$          & $-0.03(2)$	&  $(\pi \Delta)_{F}$  & $102(50)$        & $27(11)$         & $+0.15(3)$\\
$(\rho_{3} N)_P$  &  $9.1(1.5)$        & $7(1)$          & $-0.22(2)$		&$(\rho_{3} N)_P$     &  $<32$        & $<8.5$          & $+0.05(4)$\\
$(\rho_{3} N)_F$  &  $3.0(5)$      & $2.4(4)$          & $-0.128(10)$  	&$(\rho_{3} N)_F$     &  $<0.4$        & $<0.1$          & $+0.00(3)$\\
$\epsilon N $     &  $11(2)$            & $8.7(1.5)$         &  $+0.24(2)$  	&$\epsilon N $   & $192(41)$           & $51(10)$             &  $+0.20(2)$\\

\hline 
\multicolumn{4}{c}{\raisebox{-1.8ex}{\large \bf $F_{17}(1990)$**}} & \multicolumn{4}{c}{\raisebox{-1.8ex}{\large \bf $F_{17}(2200)$ new}}\\[2.5ex]
		\cline{1-4} \cline{5-8}
		
$\pi N$         & $9.4(3)$           & $1.9(4)$          & $+0.019(4)$	&$\pi N$         & $45(6)$           & $8.6(8)$       & $+0.086(7)$\\
$\eta N$        & $8.3(4.5)$            & $1.7(9)$       & $-0.018(5)$	&$\eta N$        & $22(11)$           & $4.2(2.3)$       & $+0.06(2)$ \\
$K \Lambda$     & $29(8)$           & $6.0(1)$           & $-0.034(5)$	&$K \Lambda$     & $36(9)$           & $7.0(1)$       & $-0.078(6)$ \\
	
		\hline 
		\multicolumn{4}{c}{\raisebox{-1.8ex}{\large \bf $G_{17}(2190)$****}} & \multicolumn{4}{c}{\raisebox{-1.8ex}{\large \bf $G_{19}(2250)$****}}\\[2.5ex]
		\cline{1-4} \cline{5-8}
$\pi N$         & $101(10)$           & $22.9(6)$       & $+0.229(6)$  	&$\pi N$         & $29(4)$           & $8.5(4)$         & $+0.085(4)$\\
$\eta N$        & $12(9)$           &  $2.7(2.2)$        & $+0.08(3)$  	&$\eta N$        & $<17$         & $0.07(5.0)$        & $-0.01(27)$ \\
$K \Lambda$     & $2.5(5)$           &  $0.6(1)$       & $-0.036(4)$ 	&$K \Lambda$     & $7(2)$            & $2.0(6)$         & $+0.042(6)$ \\
$(\rho_{3} N)_D$     & $<49$            & $<11$          & $-0.11(6)$  	    &             &          & \\
			
\hline \hline			  			  		
	\end{tabular}
	
\caption{\label{PartialWidths2}Below each resonance name are listed coupling constants ($\Gamma_i$), branching fractions (${\cal B}_i$), and resonant amplitudes ($\mathbf{\sqrt{xx_i}}$) for the channels listed in columns one and five. Star rating is that found in the \textit{Review of Particle Physics} (RPP)~\cite{pdg}. Table contains couplings to $D_{13}$, $D_{15}$, $F_{15}$, $F_{17}$, $G_{17}$, and $G_{19}$ resonances included in the fits.}
\end{table*}

\begin{table*}
	\begin{tabular}{@{\extracolsep{15pt}}cccc@{\extracolsep{55pt}}c @{\extracolsep{15pt}}ccc@{}}
		\hline \hline           
		\raisebox{-0.5ex}{\bf	Channel}  & \raisebox{-0.5ex}{\bf $\Gamma_i$ (MeV)} & \raisebox{-0.5ex}{ ${\cal B}_i$ }   & \raisebox{-0.5ex}{$\mathbf{\sqrt{xx_i}}$}   &  \raisebox{-0.5ex}{\bf	Channel}  & \raisebox{-0.5ex}{\bf $\Gamma_i$ (MeV)} & \raisebox{-0.5ex}{ ${\cal B}_i$ }   & \raisebox{-0.5ex}{$\mathbf{\sqrt{xx_i}}$}   \\ [1.1ex]
 \hline	
 \multicolumn{4}{c}{\raisebox{-2.0ex}{\large \bf $S_{31}(1620)$****}} & \multicolumn{4}{c}{\raisebox{-2.0ex}{\large \bf $S_{31}(1900)$***}}\\[2.5ex]
 \cline{1-4} \cline{5-8} 
		
$\pi N$         & $26(2)$           & $24(2)$          &  $+0.24(2)$	&$\pi N$         & $17(4)$            & $3.7(8)$         & $+0.037(8)$\\
$(\pi \Delta)_D$& $52(6)$            & $48(4)$         &  $-0.344(16)$	&$(\pi \Delta)_D$& $192(41)$           & $42(8)$          & $+0.12(2)$\\
$(\rho_3 N)_D$  &  $<0.05$       & $<0.04$           & $-0.003(16)$	&$(\rho_3 N)_D$  &  $83(38)$           & $18(7)$          & $-0.08(2)$\\
$\rho_1 N$      &  $29(4)$          & $27(4)$          & $+0.26(2)$	&$\rho_1 N$      &  $104(54)$          & $23(12)$          & $+0.09(2)$\\
$\pi N^*$       & $<0.02$              & $<0.02$             &  $+0.016(8)$	&$\pi N^*$       & $56(41)$            & $12(9)$          & $+0.067(25)$\\			
		\hline 
		\multicolumn{4}{c}{\raisebox{-2.0ex}{\large \bf $P_{31}(1910)$****}} & \multicolumn{4}{c}{\raisebox{-2.0ex}{\large \bf $P_{31}(2250)$ new}}\\[2.5ex]
		\cline{1-4} \cline{5-8}		
		
$\pi N$         & $34(14)$            & $13(3)$          &  $+0.13(3)$		&$\pi N$         & $45(15)$           & $14(4)$          &  $+0.14(4)$\\
$\pi N^*$       & $87(36)$            & $33(12)$             &  $-0.21(5)$		&$\pi N^*$       & $150(58)$          & $47(13)$          &  $-0.26(6)$\\
		
		\hline 
\multicolumn{4}{c}{\raisebox{-2.0ex}{\large \bf $P_{33}(1232)$****}} & \multicolumn{4}{c}{\raisebox{-2.0ex}{\large \bf $P_{33}(1600)$****}}\\[2.5ex]
		\cline{1-4} \cline{5-8}
$\pi N$         & $110.2(8)$      & $99.39(1)$      &  $+0.994(1)$	&$\pi N$         & $34(8)$      & $10.7(1.9)$      &  $+0.107(19)$\\
$(\pi \Delta)_P$ & $0.0(0)$       & $0.0(0)$          & $+0.00(1)$		&$(\pi \Delta)_P$ & $206(28)$       & $64(6)$          & $+0.26(2)$\\
$\pi N^*$       & $0.0(0)$        & $0.0(0)$             & $+0.00(1)$		&$\pi N^*$       & $70(18)$       & $22(5)$             &  $+0.15(2)$\\
		
		\hline 
\multicolumn{4}{c}{\raisebox{-2.0ex}{\large \bf $P_{33}(1920)$***}} & \multicolumn{4}{c}{\raisebox{-2.0ex}{\large \bf $D_{33}(1700)$****}}\\[2.5ex]
		\cline{1-4} \cline{5-8}
$\pi N$         & $53(25)$      & $10.5(3.0)$      &  $+0.10(3)$		&$\pi N$         & $34(4)$          & $15(2)$        &  $+0.15(2)$\\
$(\pi \Delta)_P$ & $<8$       & $<1.6$          & $-0.017(39)$		&$(\pi \Delta)_S$& $112(13)$         & $49(5)$        &  $+0.27(2)$\\
$\pi N^*$       & $392(94)$        & $77(9)$             & $+0.28(4)$	&$(\pi \Delta)_D$& $17(7)$          & $7.6(3)$       &  $-0.11(2)$\\
 &          &            & 		&$(\rho_3 N)_S$  &  $62(14)$         & $27(5)$        &  $+0.20(2)$\\
\hline
\multicolumn{4}{c}{\raisebox{-2.0ex}{\large \bf $D_{33}(1940)$**}} & \multicolumn{4}{c}{\raisebox{-2.0ex}{\large \bf $D_{35}(1930)$***}}\\[2.5ex]
		\cline{1-4} \cline{5-8}
	
$\pi N$         & $62(14)$           & $16(4)$          & $+0.16(4)$		&$\pi N$         & $47(13)$          & $9.5(1)$          &  $+0.095(10)$\\
$(\pi \Delta)_S$& $<3.6$         & $<0.9$         & $+0.018(32)$	& &        &          &  \\
$(\pi \Delta)_D$& $<25$         & $<6.3$         & $-0.068(38)$ &  &  &  &\\
$(\rho_3 N)_S$  &  $321(47)$         & $80(5)$     & $+0.35(4)$  &  &  &  &\\			
			
\hline
\multicolumn{4}{c}{\raisebox{-2.0ex}{\large \bf $F_{35}(1905)$****}} & \multicolumn{4}{c}{\raisebox{-2.0ex}{\large \bf $F_{37}(1950)$****}}\\[2.5ex]
\cline{1-4} \cline{5-8}	
		
$\pi N$         & $50(5)$         & $17(1)$         &  $+0.17(1)$		&$\pi N$         & $92(6)$           & $38(2)$           & $+0.383(15)$\\
$(\pi \Delta)_P$& $24(15)$          & $8.4(5)$           &  $+0.12(4)$		& &            &            & \\
$(\pi \Delta)_F$& $140(27)$          & $49(9)$         &  $+0.29(3)$		& &          &            & \\
$(\rho_3 N)_P$  &  $74(27)$          & $26(9)$         &  $+0.21(4)$	&     &            &         & \\

\hline \hline

\end{tabular}

\caption{\label{PartialWidths3}Below each resonance name are listed coupling constants ($\Gamma_i$), branching fractions (${\cal B}_i$), and resonant amplitudes ($\mathbf{\sqrt{xx_i}}$) for the channels listed in columns one and five. Star rating is that found in the \textit{Review of Particle Physics} (RPP)~\cite{pdg}. Table contains couplings to $S_{31}$, $P_{31}$, $P_{33}$, $D_{33}$, $D_{35}$, $F_{35}$, $F_{37}$ resonances included in the fits.}
\end{table*}
\onecolumngrid
\newcommand{\ArgandScale}{.37}
\newcommand{\trimA}{30mm}
\newcommand{\trimB}{42mm}
\newcommand{\trimC}{20mm}
\newcommand{\trimD}{30mm}

\begin{minipage}[p]{\textwidth}
	\includegraphics[trim={\trimA} {\trimB} {\trimC} {\trimD},clip=true]{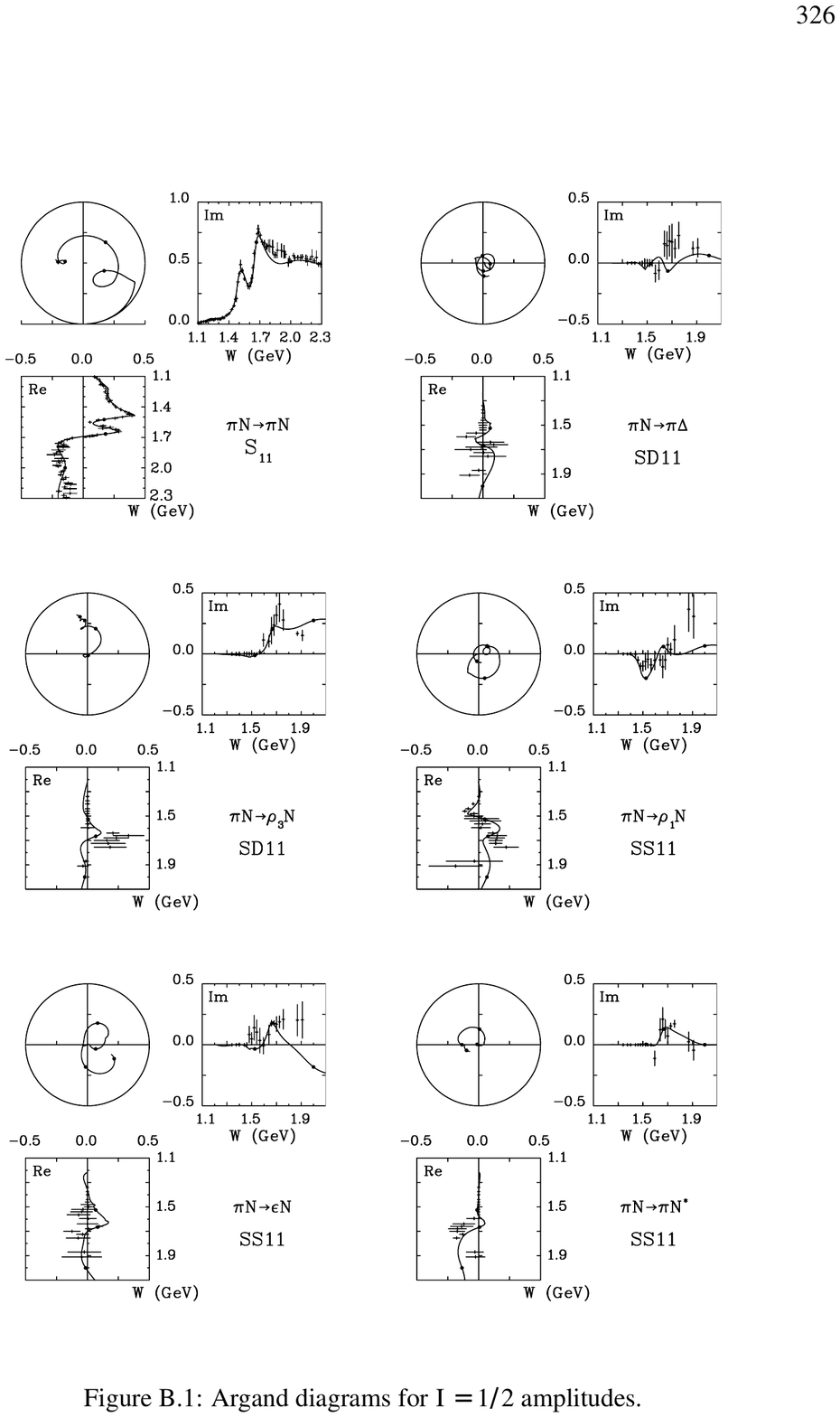}
	\captionof{figure}{Argand diagrams for the $I = 1/2$ amplitudes.}
	\label{Argand01}
\end{minipage}
\newpage

\begin{minipage}[p]{\textwidth}
	\includegraphics[trim={\trimA} {\trimB} {\trimC} {\trimD},clip=true]{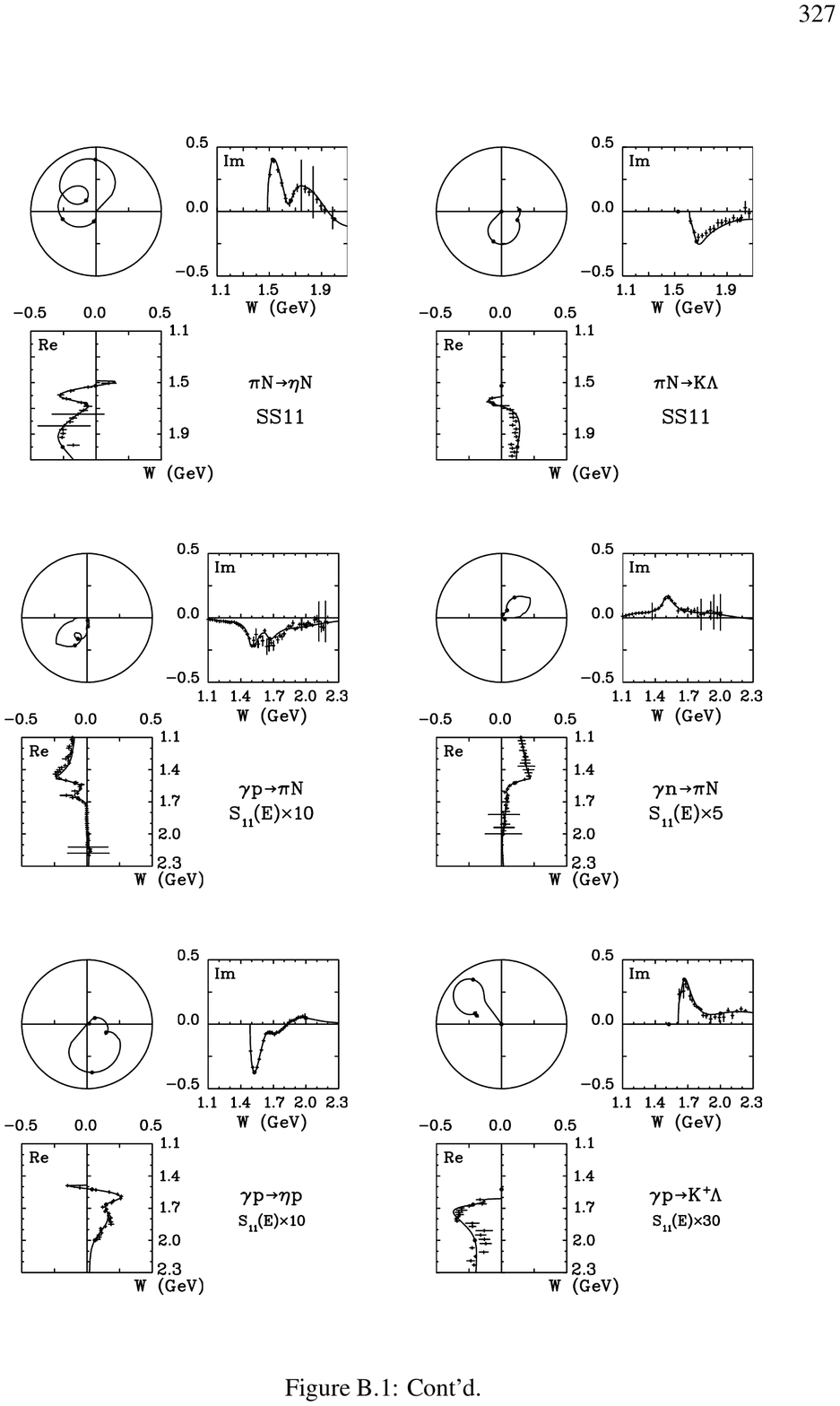}
	\captionof{figure}{Argand diagrams for the $I = 1/2$ amplitudes.}
	\label{Argand02}
\end{minipage}
\newpage
\begin{minipage}[p]{\textwidth}
	\includegraphics[trim={\trimA} {\trimB} {\trimC} {\trimD},clip=true]{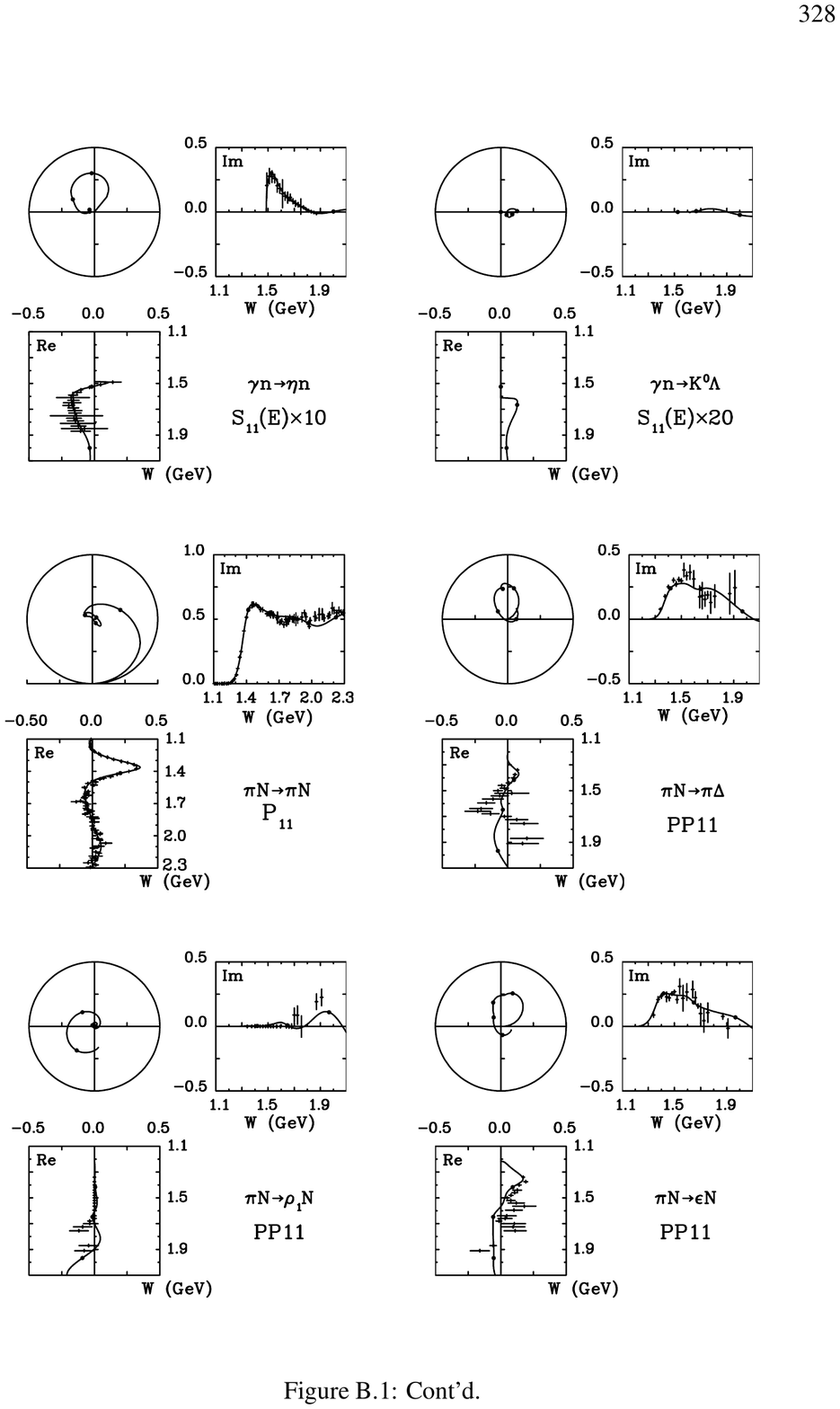}
	\captionof{figure}{Argand diagrams for the $I = 1/2$ amplitudes.}
	\label{Argand03}
\end{minipage}
\newpage
\begin{minipage}[p]{\textwidth}
	\includegraphics[trim={\trimA} {\trimB} {\trimC} {\trimD},clip=true]{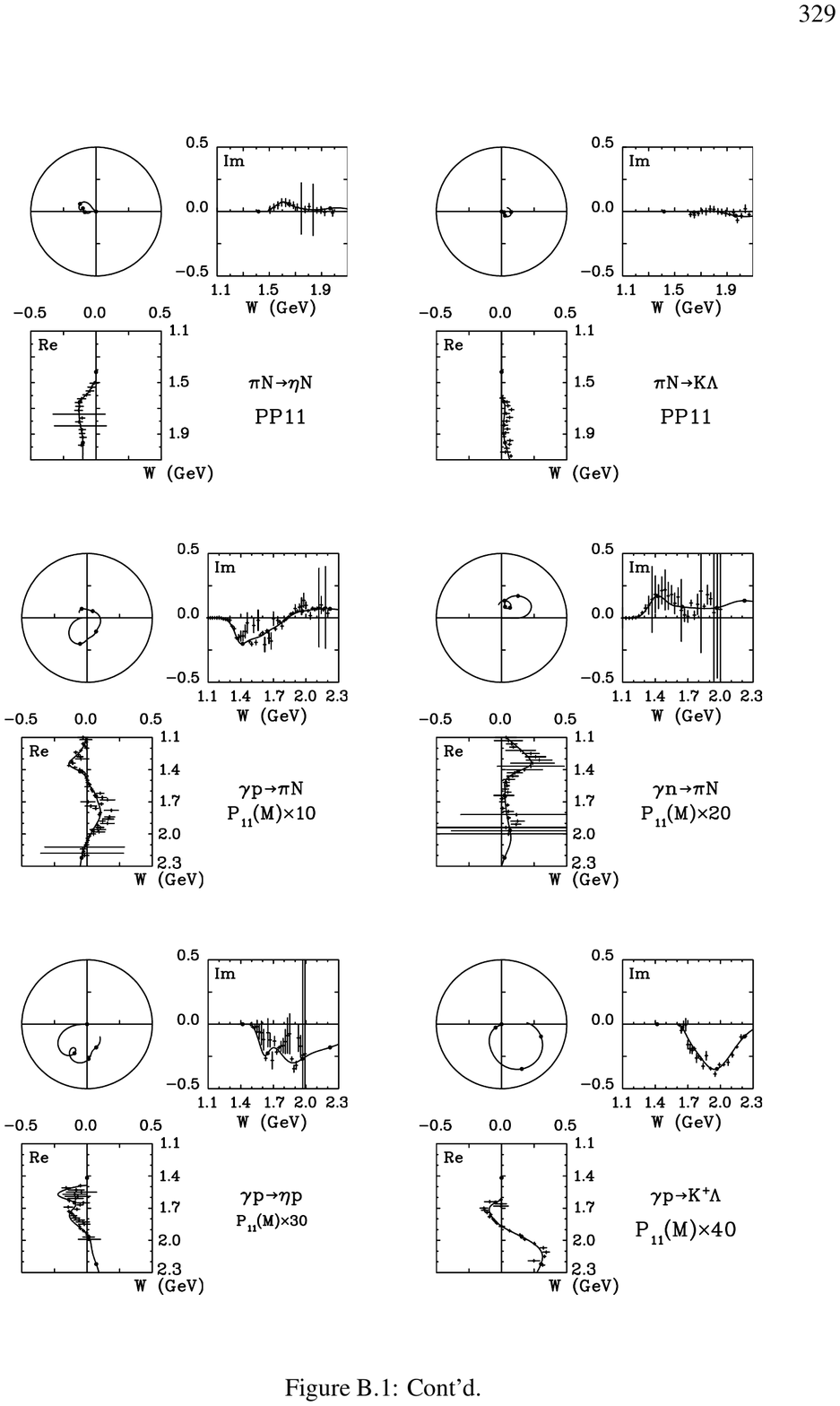}
	\captionof{figure}{Argand diagrams for the $I = 1/2$ amplitudes.}
	\label{Argand04}
\end{minipage}
\newpage
\begin{minipage}[p]{\textwidth}
	\includegraphics[trim={\trimA} {\trimB} {\trimC} {\trimD},clip=true]{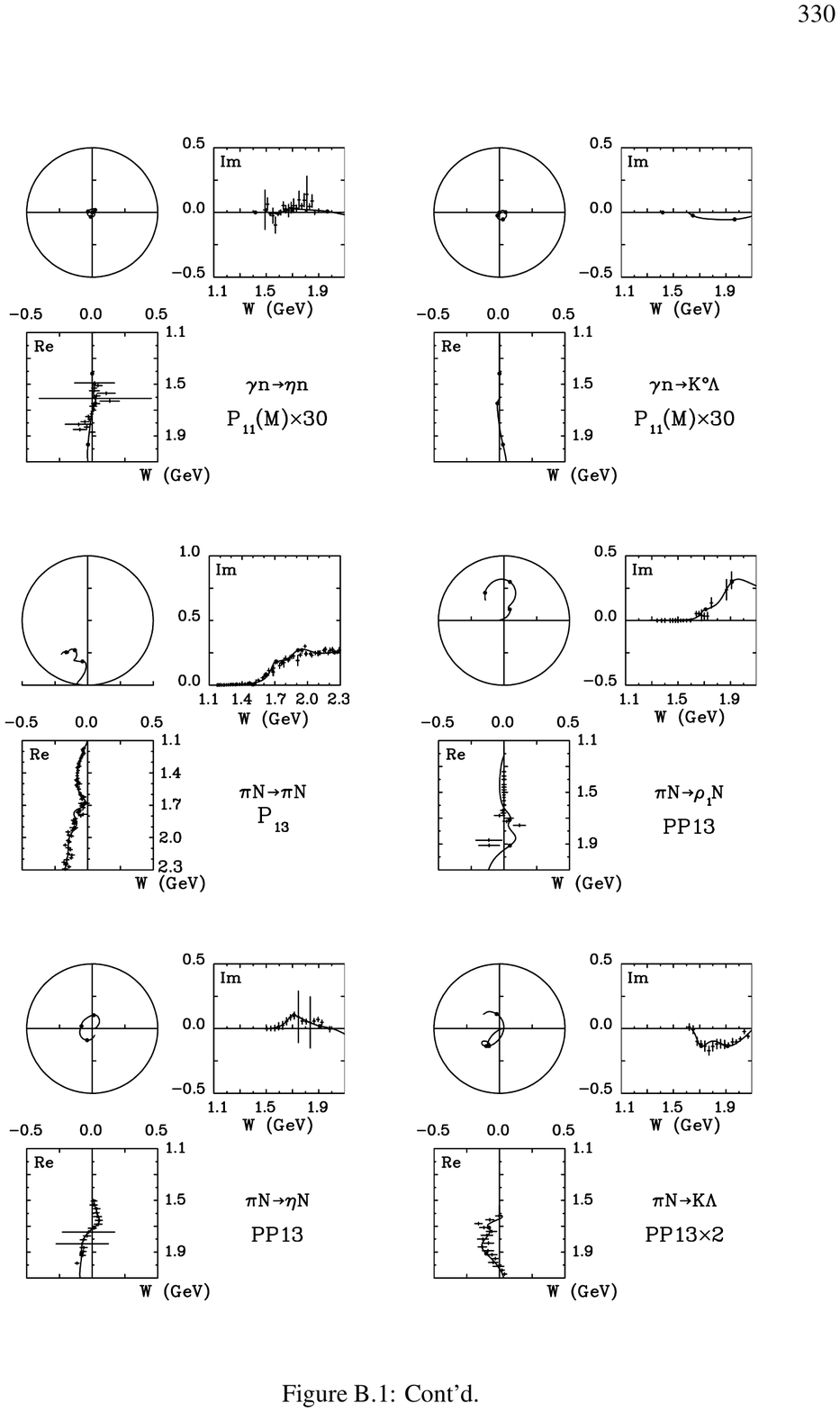}
	\captionof{figure}{Argand diagrams for the $I = 1/2$ amplitudes.}
	\label{Argand05}
\end{minipage}
\newpage
\begin{minipage}[p]{\textwidth}
	\includegraphics[trim={\trimA} {\trimB} {\trimC} {\trimD},clip=true]{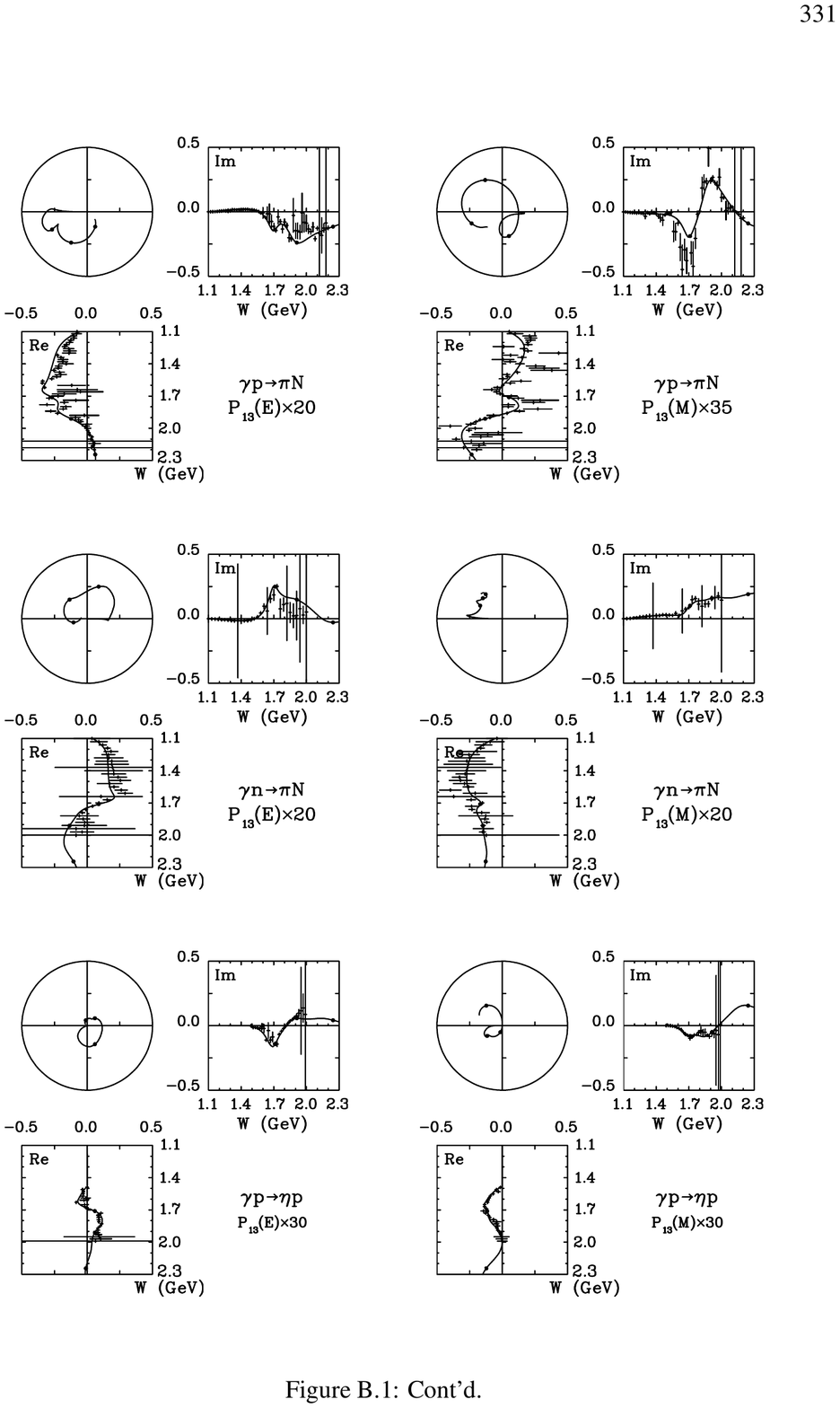}
	\captionof{figure}{Argand diagrams for the $I = 1/2$ amplitudes.}
	\label{Argand06}
\end{minipage}
\newpage
\begin{minipage}[p]{\textwidth}
	\includegraphics[trim={\trimA} {\trimB} {\trimC} {\trimD},clip=true]{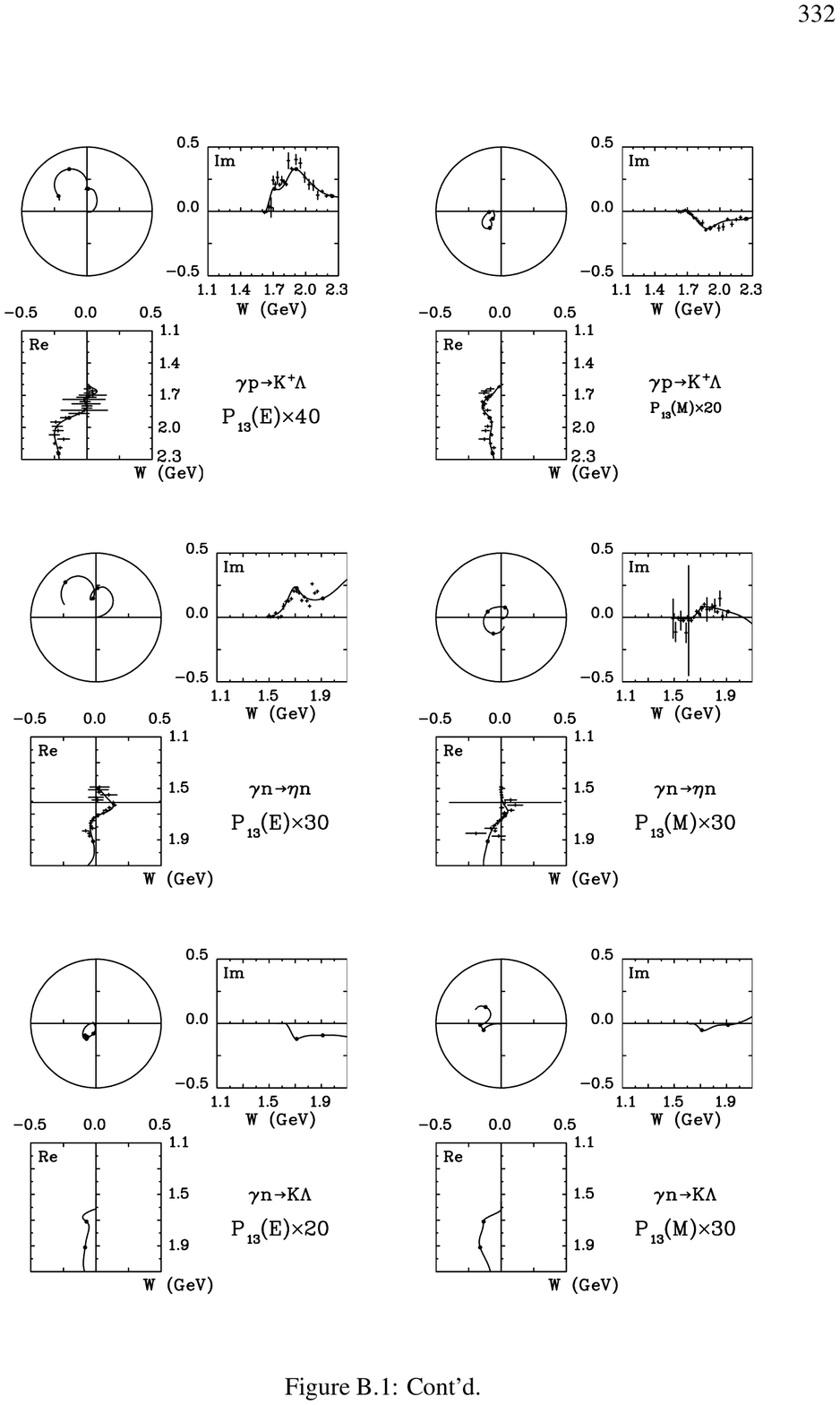}
	\captionof{figure}{Argand diagrams for the $I = 1/2$ amplitudes.}
	\label{Argand07}
\end{minipage}
\newpage
\begin{minipage}[p]{\textwidth}
	\includegraphics[trim={\trimA} {\trimB} {\trimC} {\trimD},clip=true]{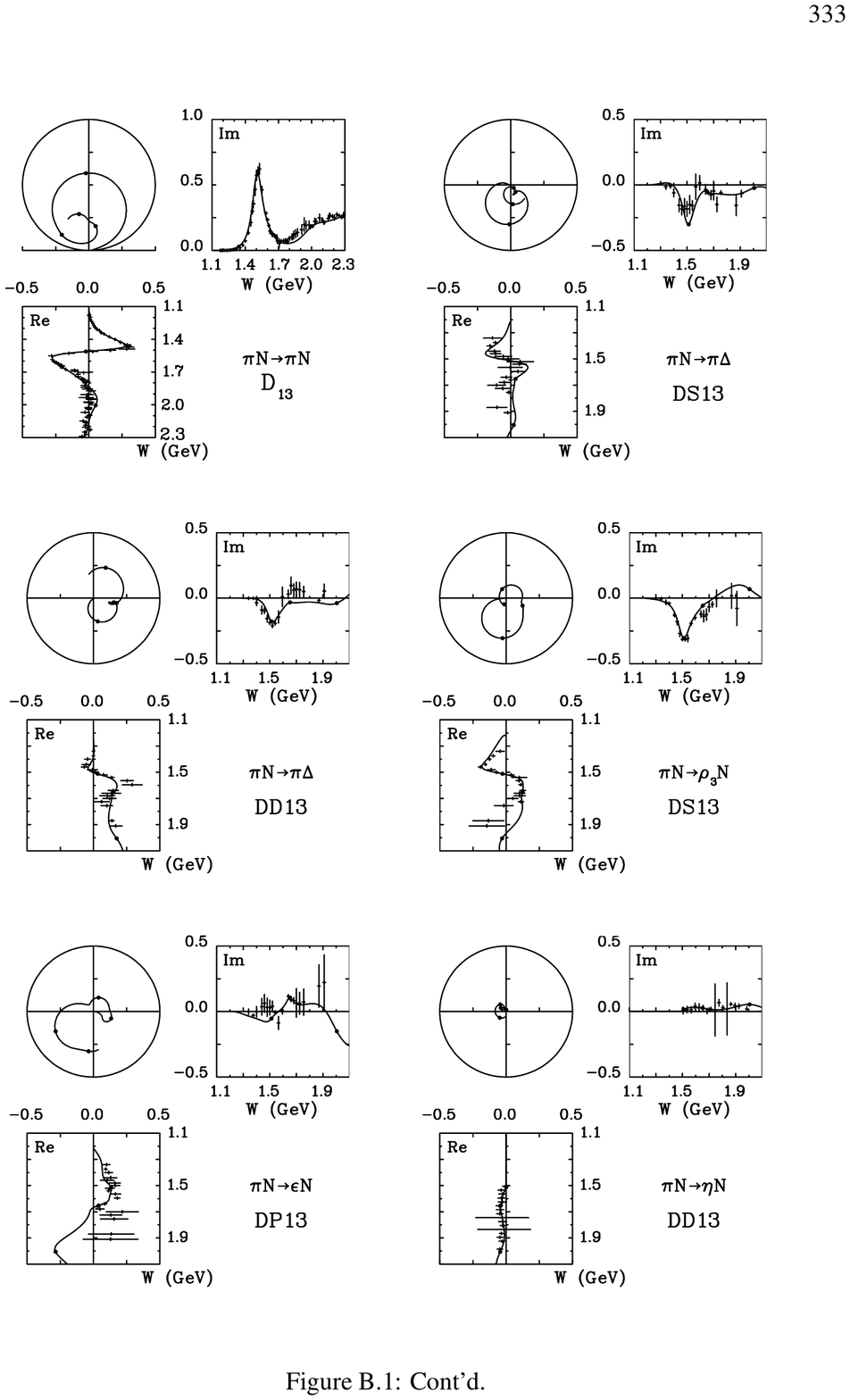}
	\captionof{figure}{Argand diagrams for the $I = 1/2$ amplitudes.}
	\label{Argand08}
\end{minipage}
\newpage
\begin{minipage}[p]{\textwidth}
	\includegraphics[trim={\trimA} {\trimB} {\trimC} {\trimD},clip=true]{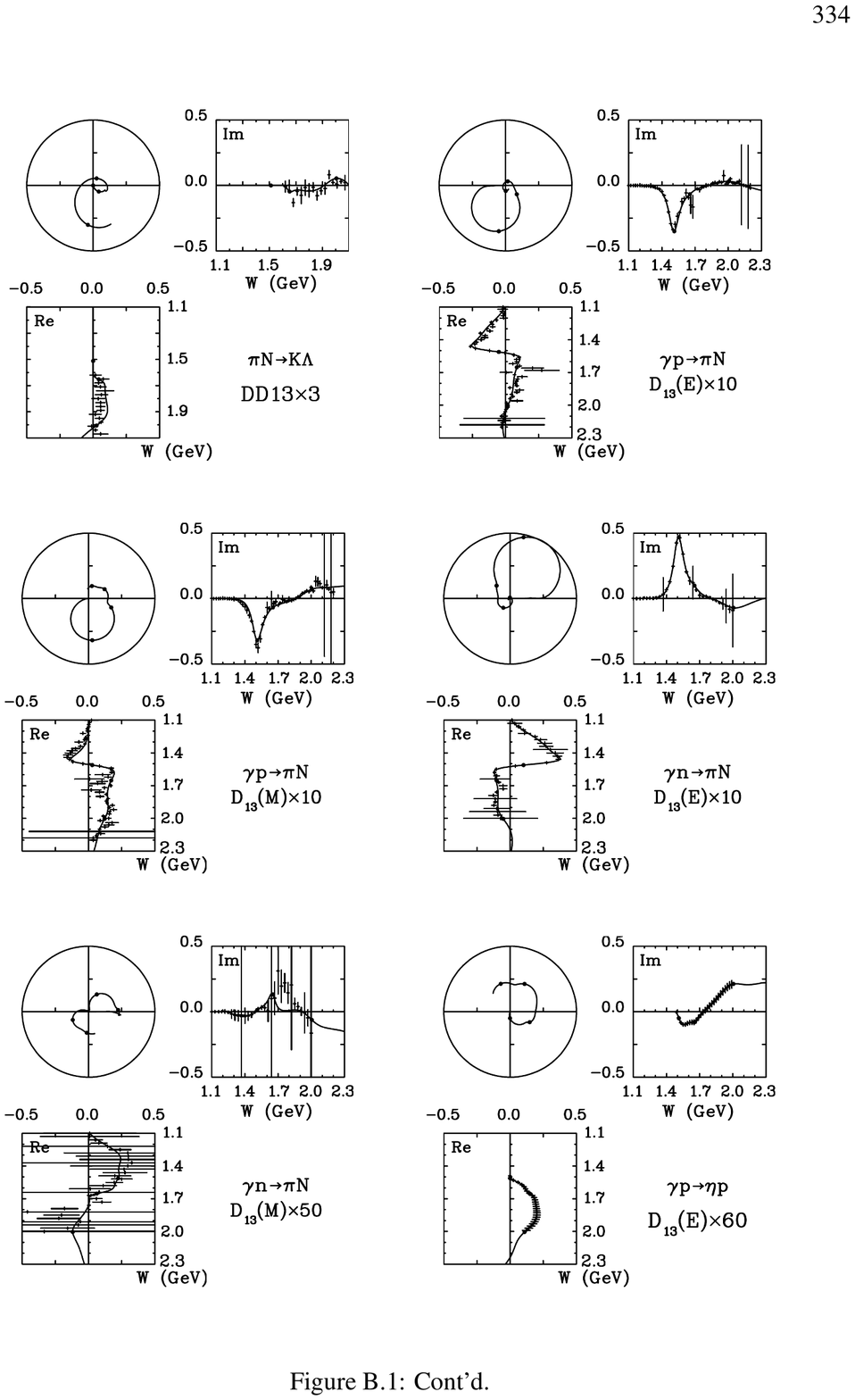}
	\captionof{figure}{Argand diagrams for the $I = 1/2$ amplitudes.}
	\label{Argand09}
\end{minipage}
\newpage
\begin{minipage}[p]{\textwidth}
	\includegraphics[trim={\trimA} {\trimB} {\trimC} {\trimD},clip=true]{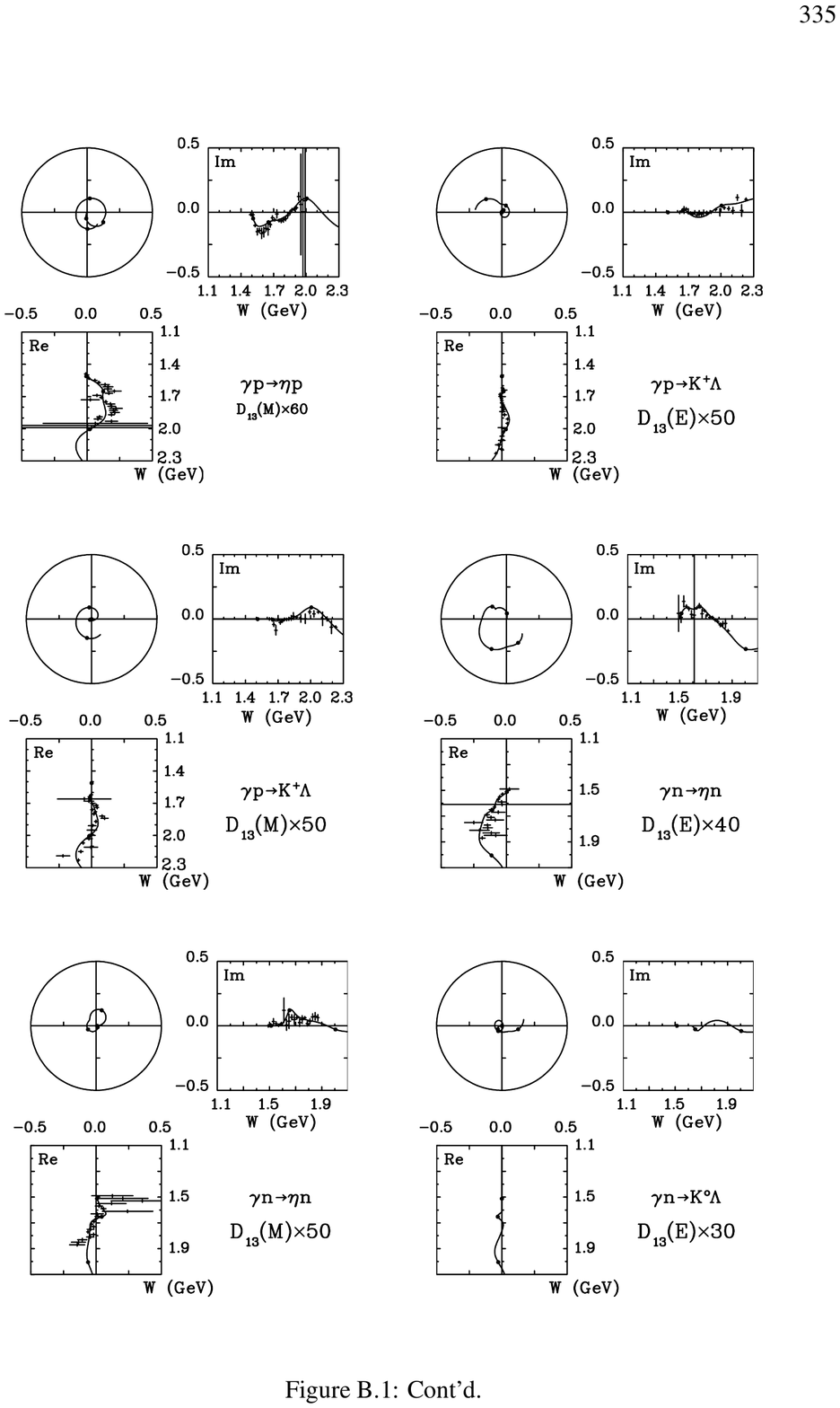}
	\captionof{figure}{Argand diagrams for the $I = 1/2$ amplitudes.}
	\label{Argand10}
\end{minipage}
\newpage
\begin{minipage}[p]{\textwidth}
	\includegraphics[trim={\trimA} {\trimB} {\trimC} {\trimD},clip=true]{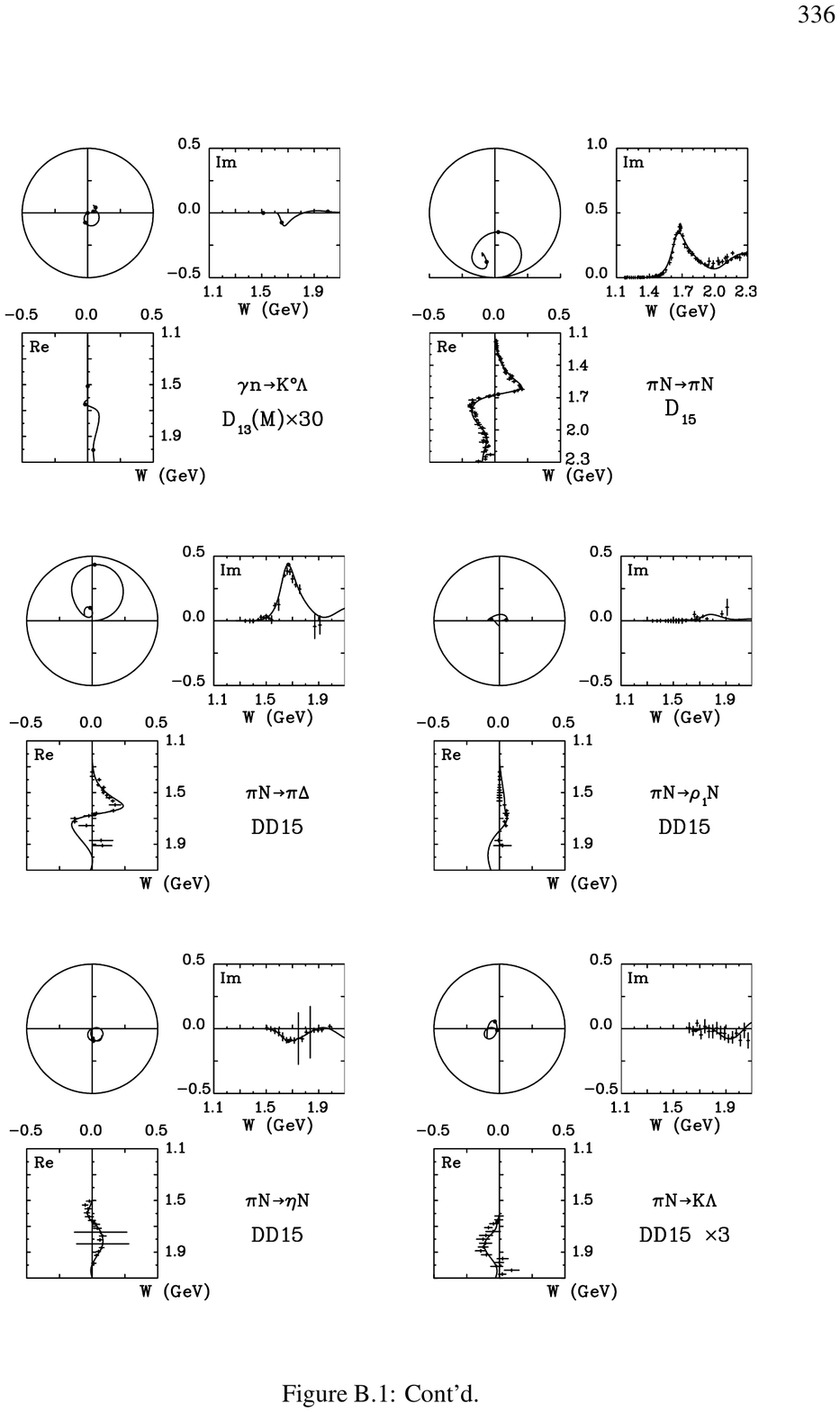}
	\captionof{figure}{Argand diagrams for the $I = 1/2$ amplitudes.}
	\label{Argand11}
\end{minipage}
\newpage
\begin{minipage}[p]{\textwidth}
	\includegraphics[trim={\trimA} {\trimB} {\trimC} {\trimD},clip=true]{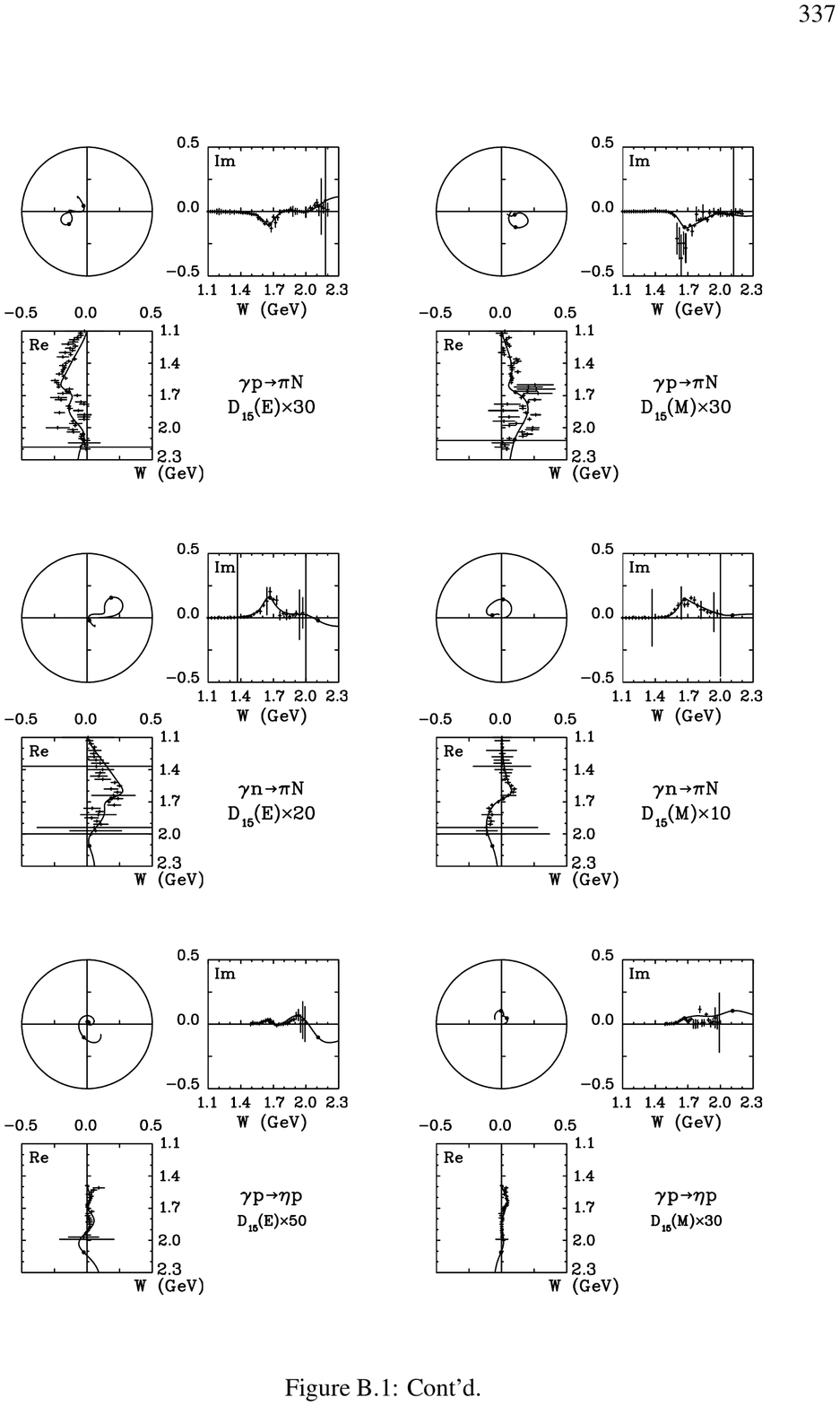}
	\captionof{figure}{Argand diagrams for the $I = 1/2$ amplitudes.}
	\label{Argand12}
\end{minipage}
\newpage
\begin{minipage}[p]{\textwidth}
	\includegraphics[trim={\trimA} {\trimB} {\trimC} {\trimD},clip=true]{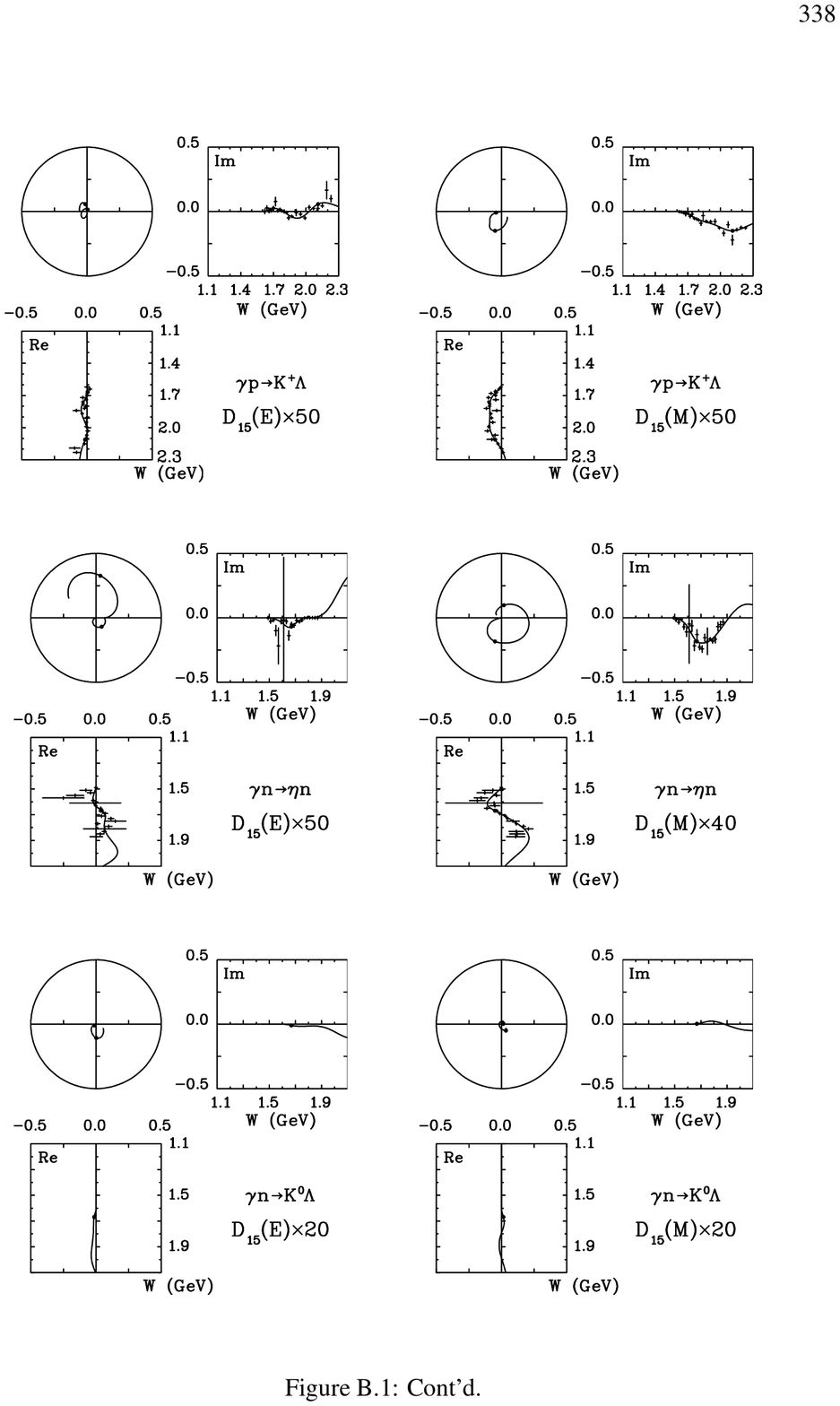}
	\captionof{figure}{Argand diagrams for the $I = 1/2$ amplitudes.}
	\label{Argand13}
\end{minipage}
\newpage
\begin{minipage}[p]{\textwidth}
	\includegraphics[trim={\trimA} {\trimB} {\trimC} {\trimD},clip=true]{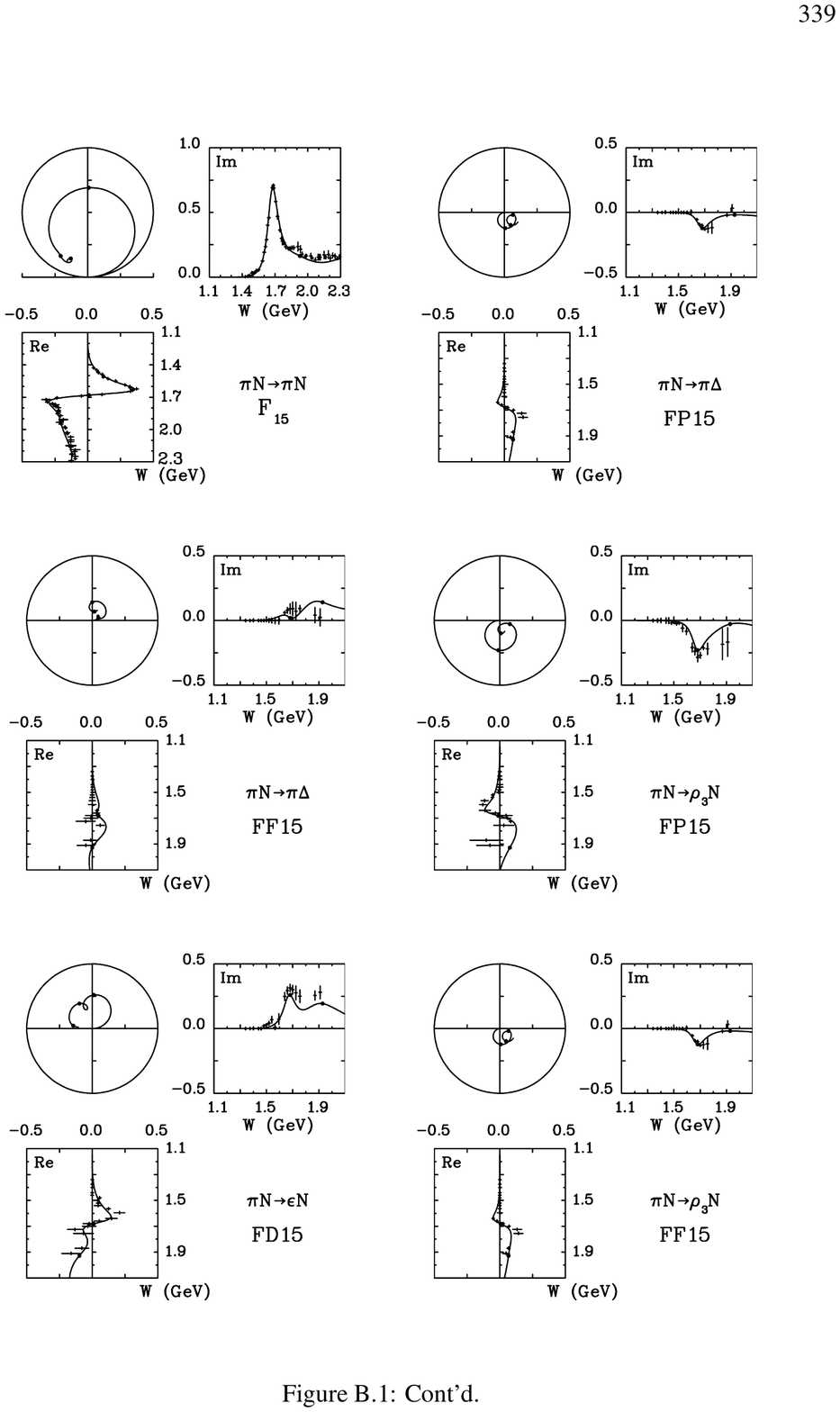}
	\captionof{figure}{Argand diagrams for the $I = 1/2$ amplitudes.}
	\label{Argand14}
\end{minipage}
\newpage
\begin{minipage}[p]{\textwidth}
	\includegraphics[trim={\trimA} {\trimB} {\trimC} {\trimD},clip=true]{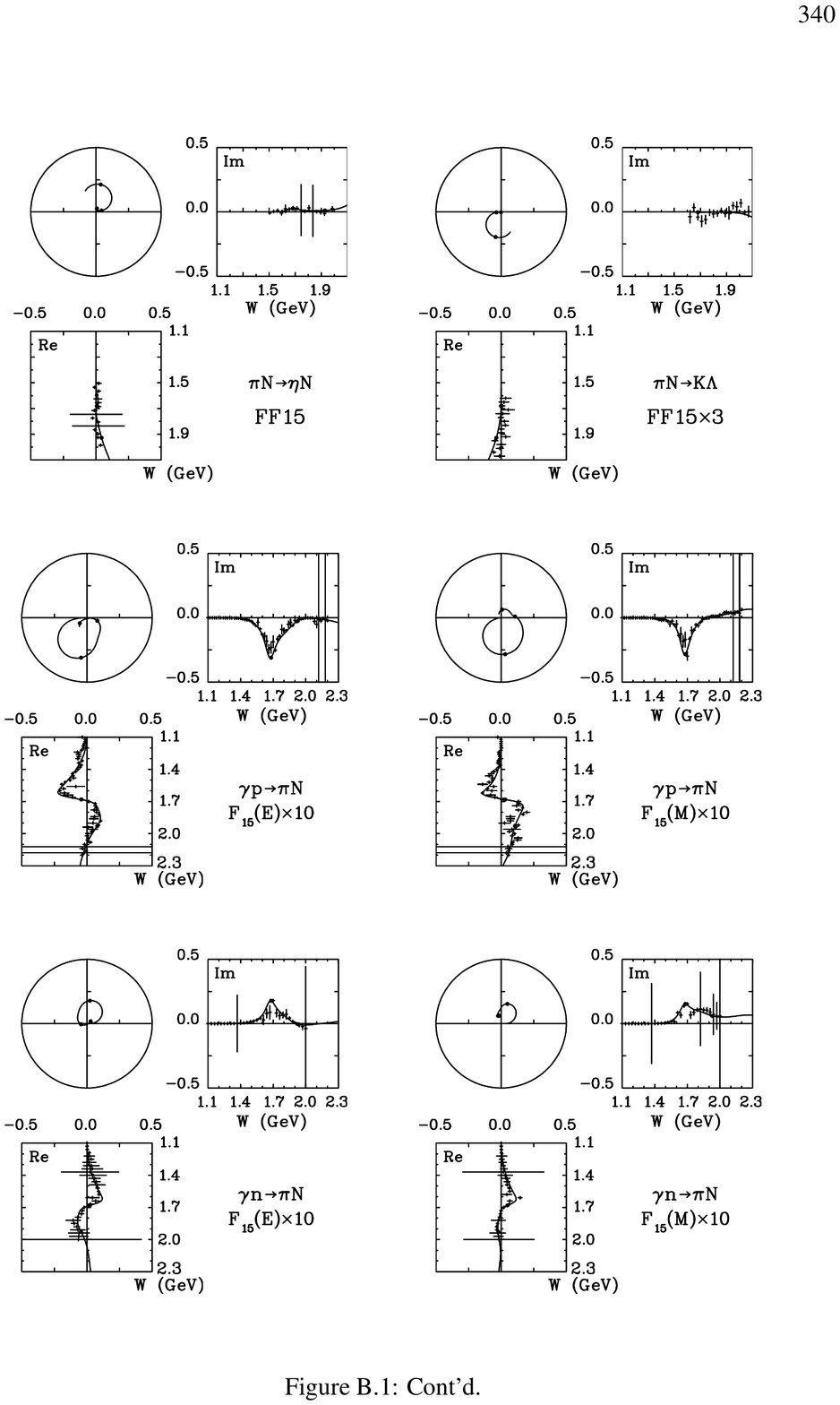}
	\captionof{figure}{Argand diagrams for the $I = 1/2$ amplitudes.}
	\label{Argand15}
\end{minipage}
\newpage
\begin{minipage}[p]{\textwidth}
	\includegraphics[trim={\trimA} {\trimB} {\trimC} {\trimD},clip=true]{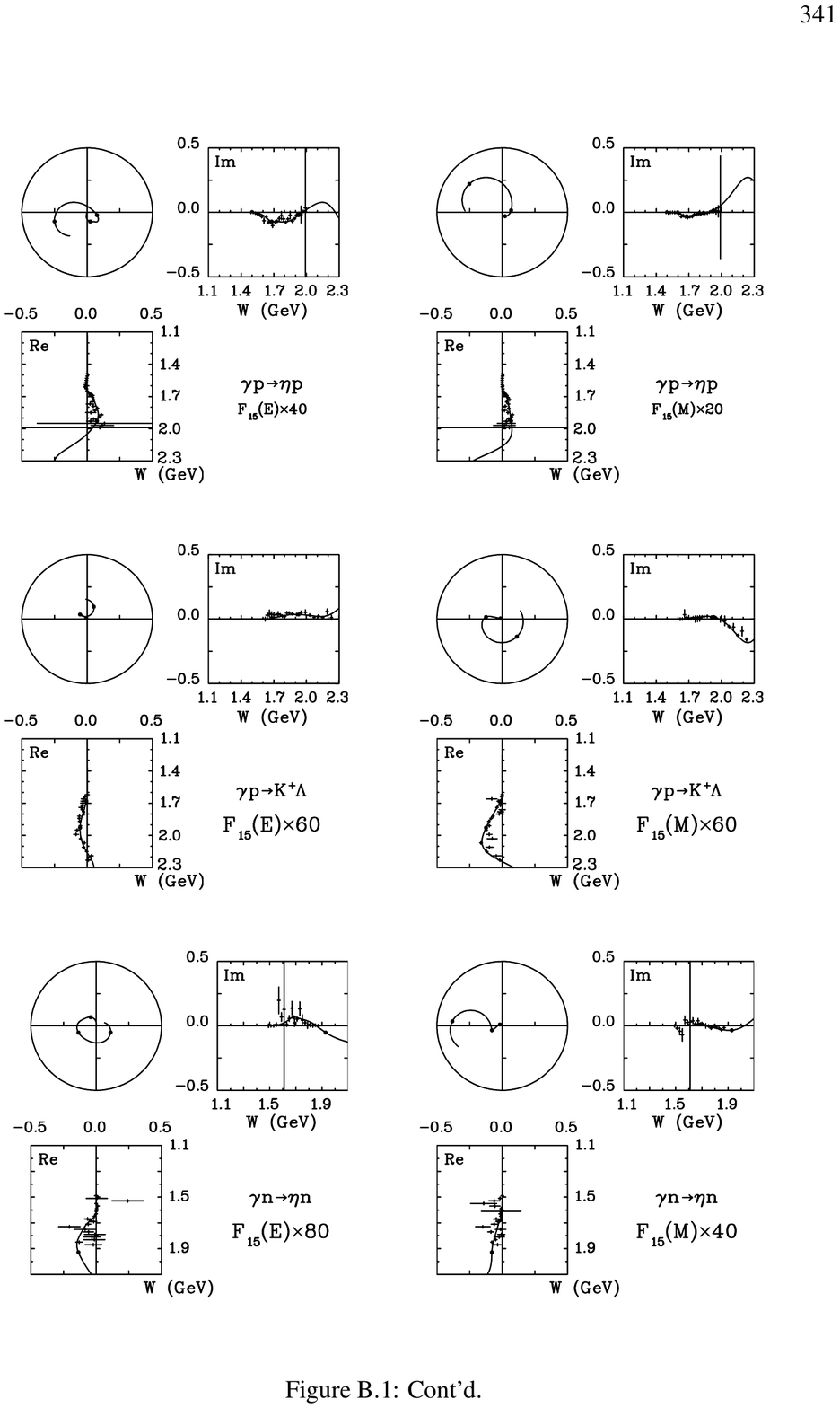}
	\captionof{figure}{Argand diagrams for the $I = 1/2$ amplitudes.}
	\label{Argand16}
\end{minipage}
\newpage
\begin{minipage}[p]{\textwidth}
	\includegraphics[trim={\trimA} {\trimB} {\trimC} {\trimD},clip=true]{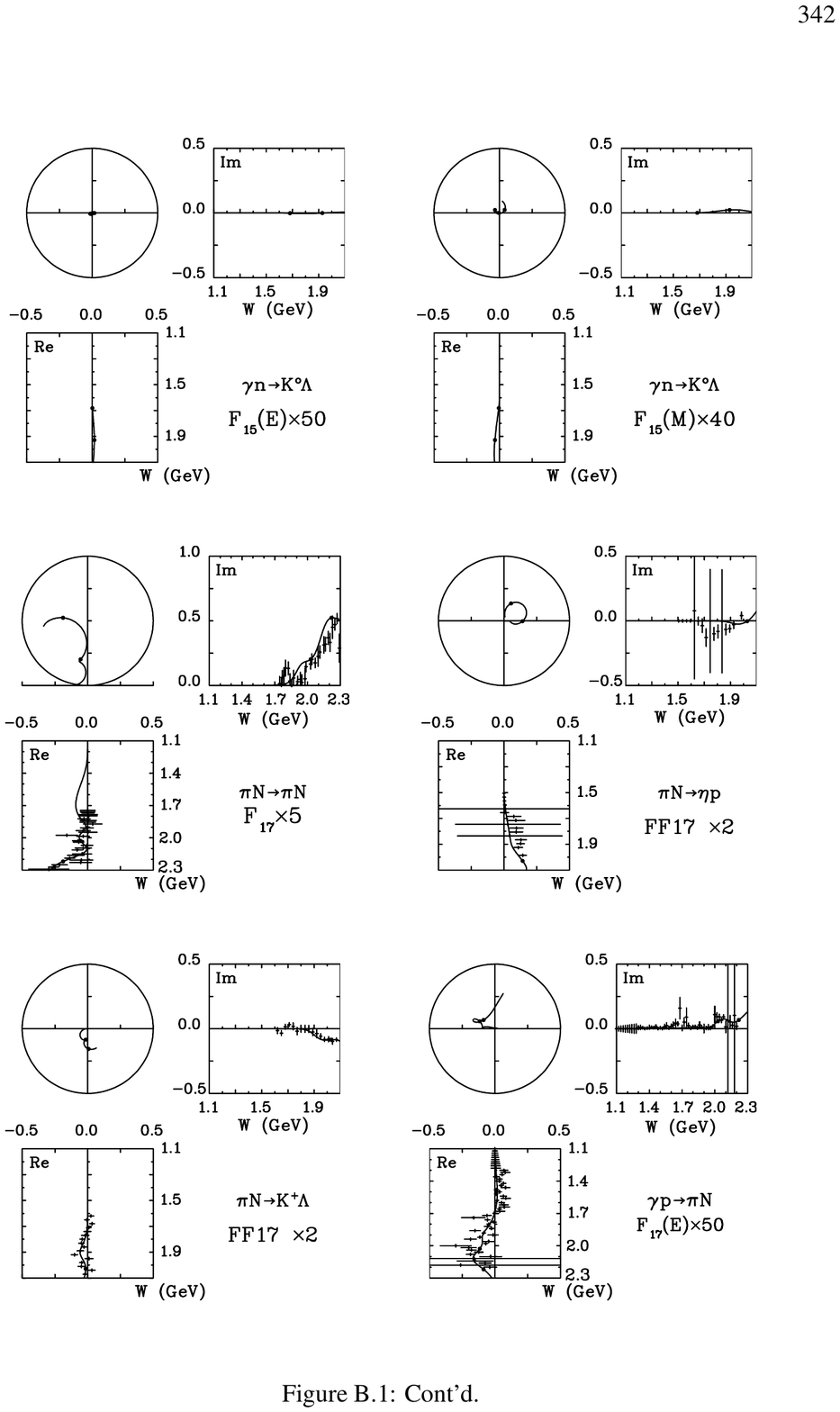}
	\captionof{figure}{Argand diagrams for the $I = 1/2$ amplitudes.}
	\label{Argand17}
\end{minipage}
\newpage
\begin{minipage}[p]{\textwidth}
	\includegraphics[trim={\trimA} {\trimB} {\trimC} {\trimD},clip=true]{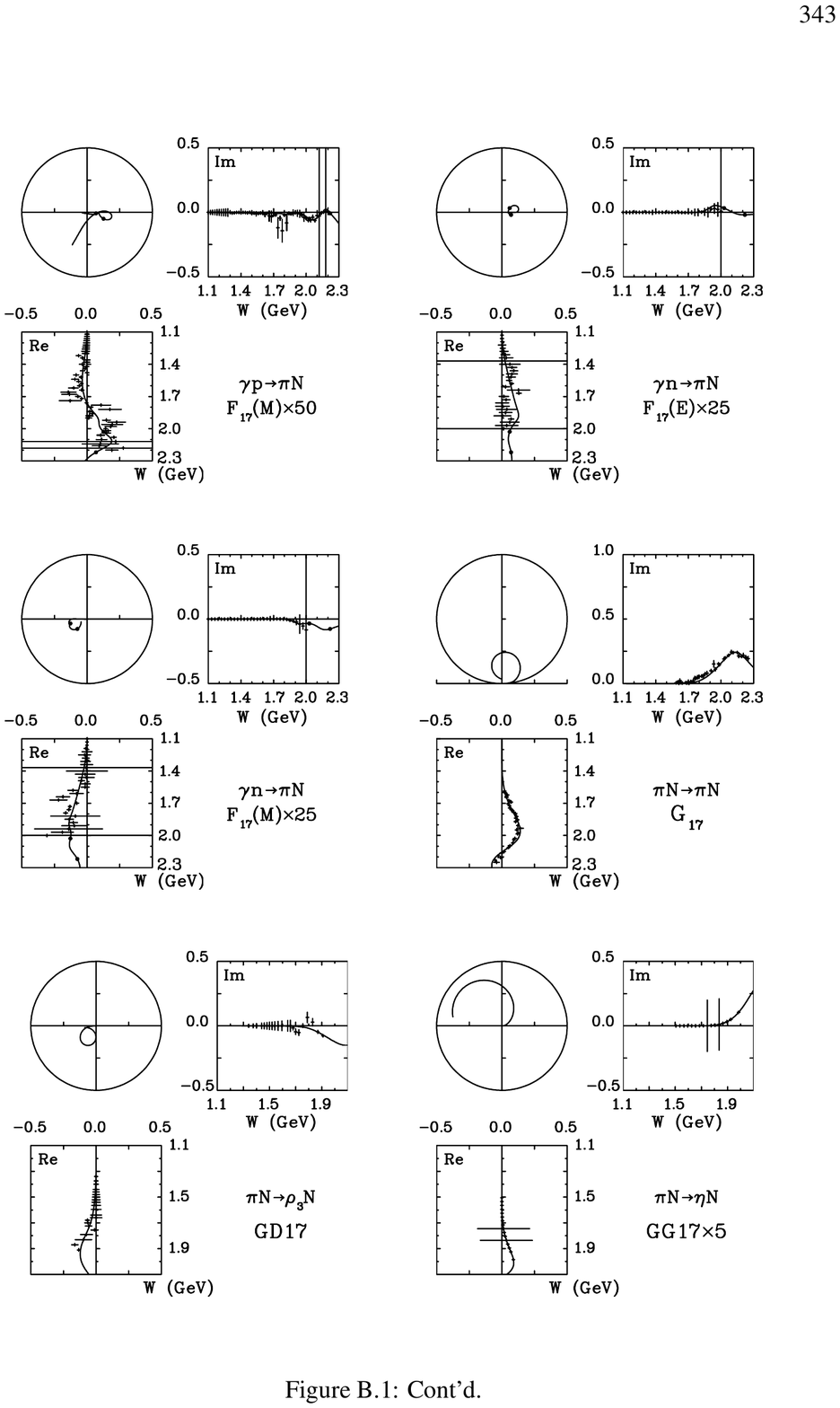}
	\captionof{figure}{Argand diagrams for the $I = 1/2$ amplitudes.}
	\label{Argand18}
\end{minipage}
\newpage
\begin{minipage}[p]{\textwidth}
	\includegraphics[trim={\trimA} {\trimB} {\trimC} {\trimD},clip=true]{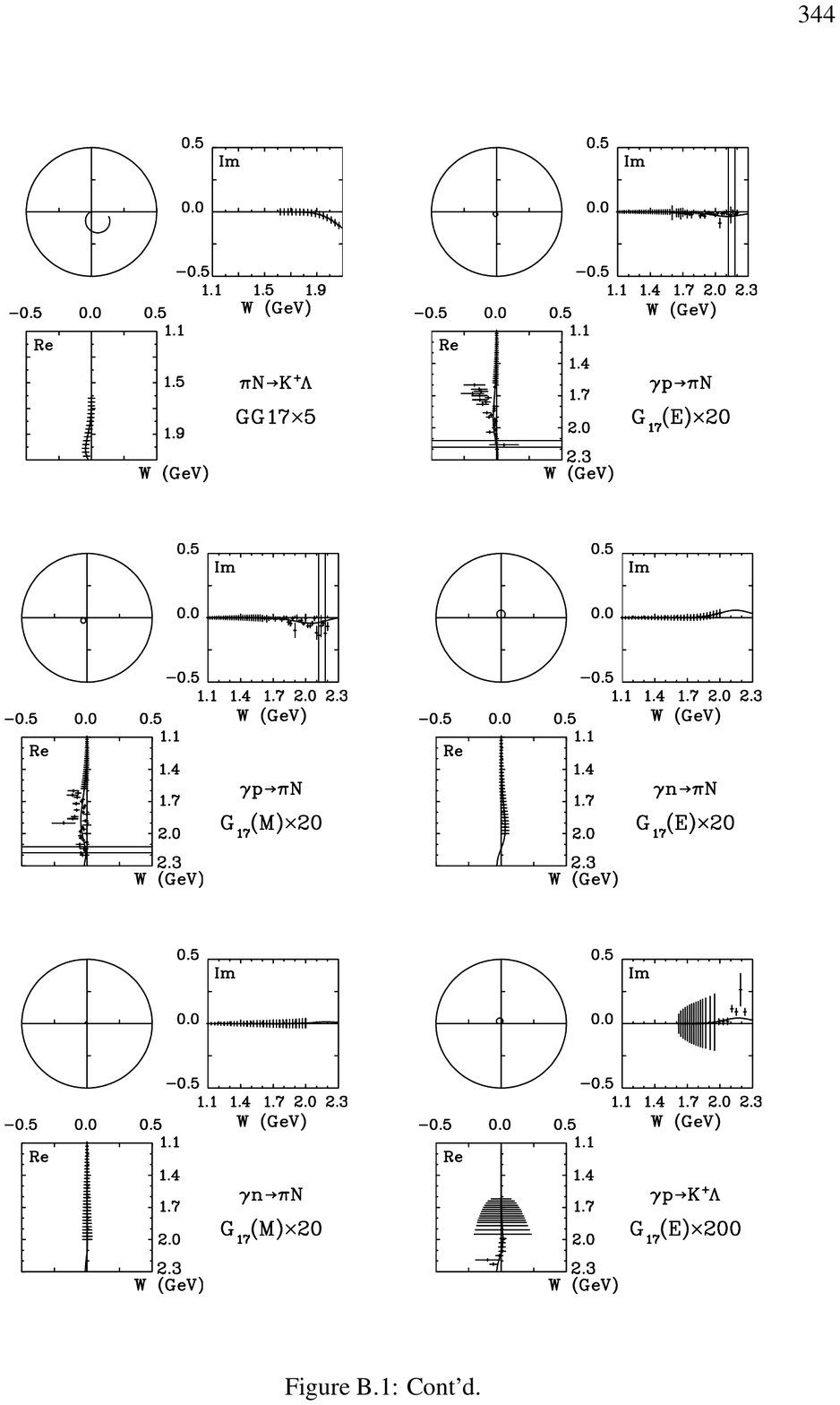}
	\captionof{figure}{Argand diagrams for the $I = 1/2$ amplitudes.}
	\label{Argand19}
\end{minipage}
\newpage
\begin{minipage}[p]{\textwidth}
	\includegraphics[trim={\trimA} {\trimB} {\trimC} {\trimD},clip=true]{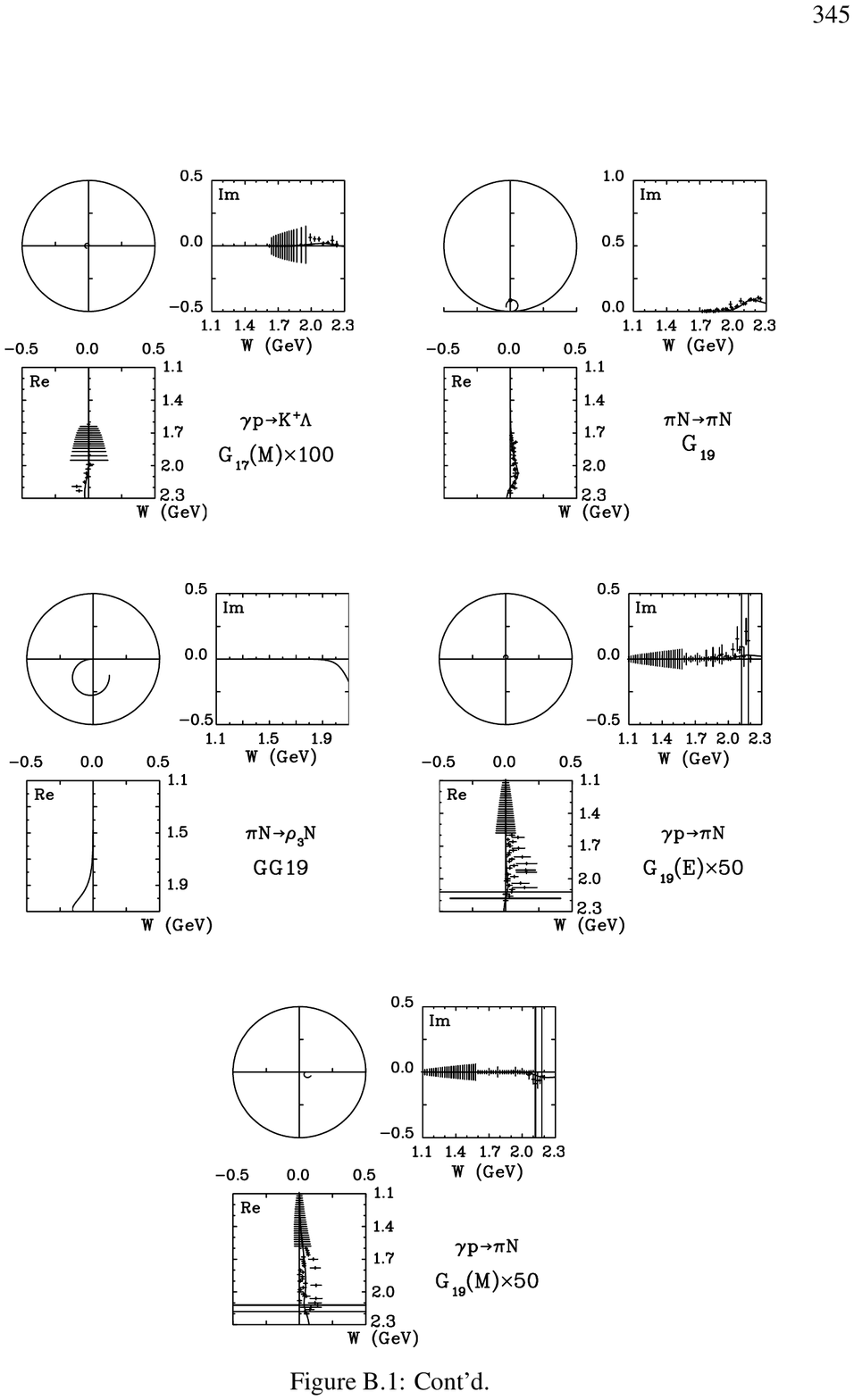}
	\captionof{figure}{Argand diagrams for the $I = 1/2$ amplitudes.}
	\label{Argand20}
\end{minipage}
\newpage
\begin{minipage}[p]{\textwidth}
	\includegraphics[trim={\trimA} {\trimB} {\trimC} {\trimD},clip=true]{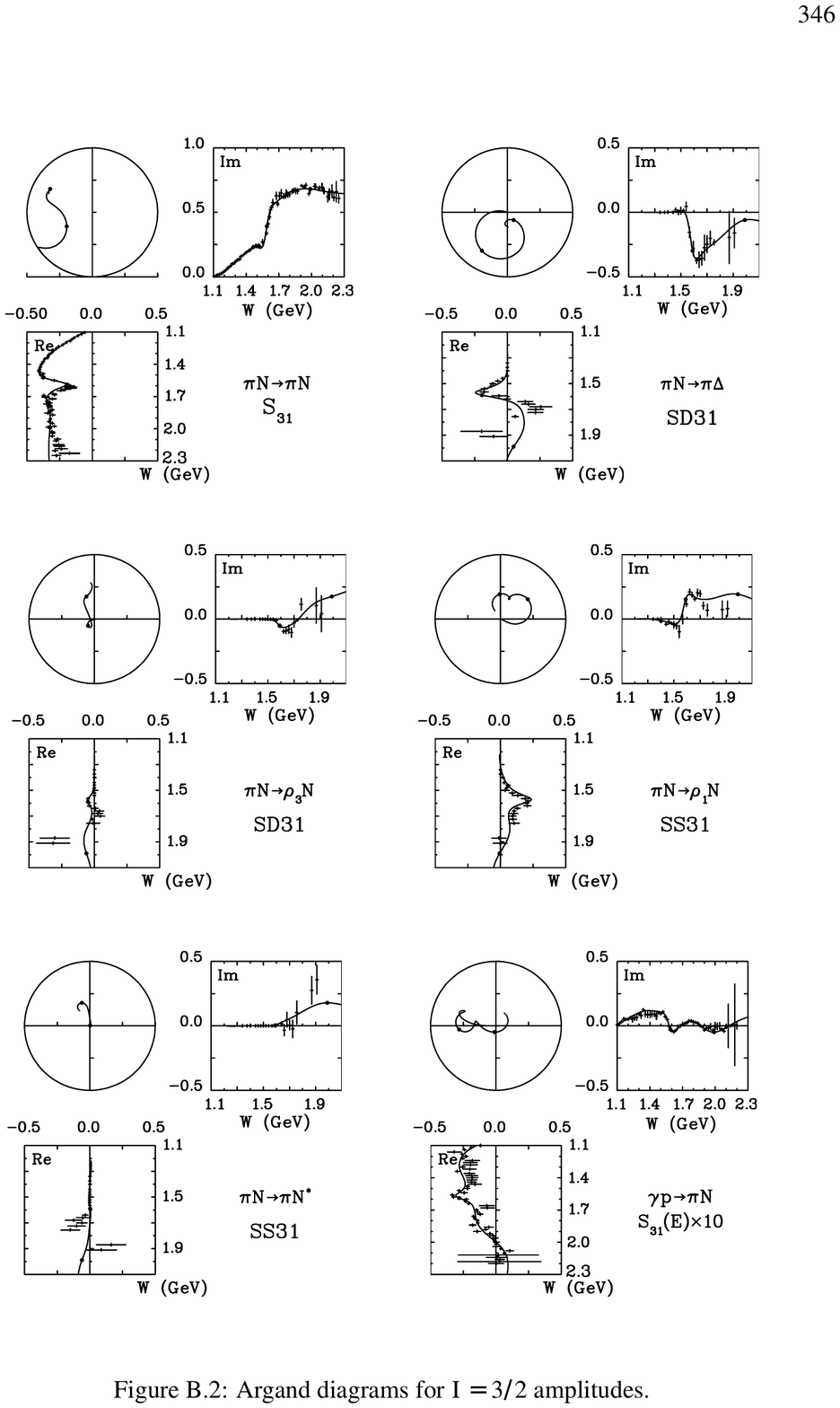}
	\captionof{figure}{Argand diagrams for the $I = 3/2$ amplitudes.}
	\label{Argand21}
\end{minipage}
\newpage
\begin{minipage}[p]{\textwidth}
	\includegraphics[trim={\trimA} {\trimB} {\trimC} {\trimD},clip=true]{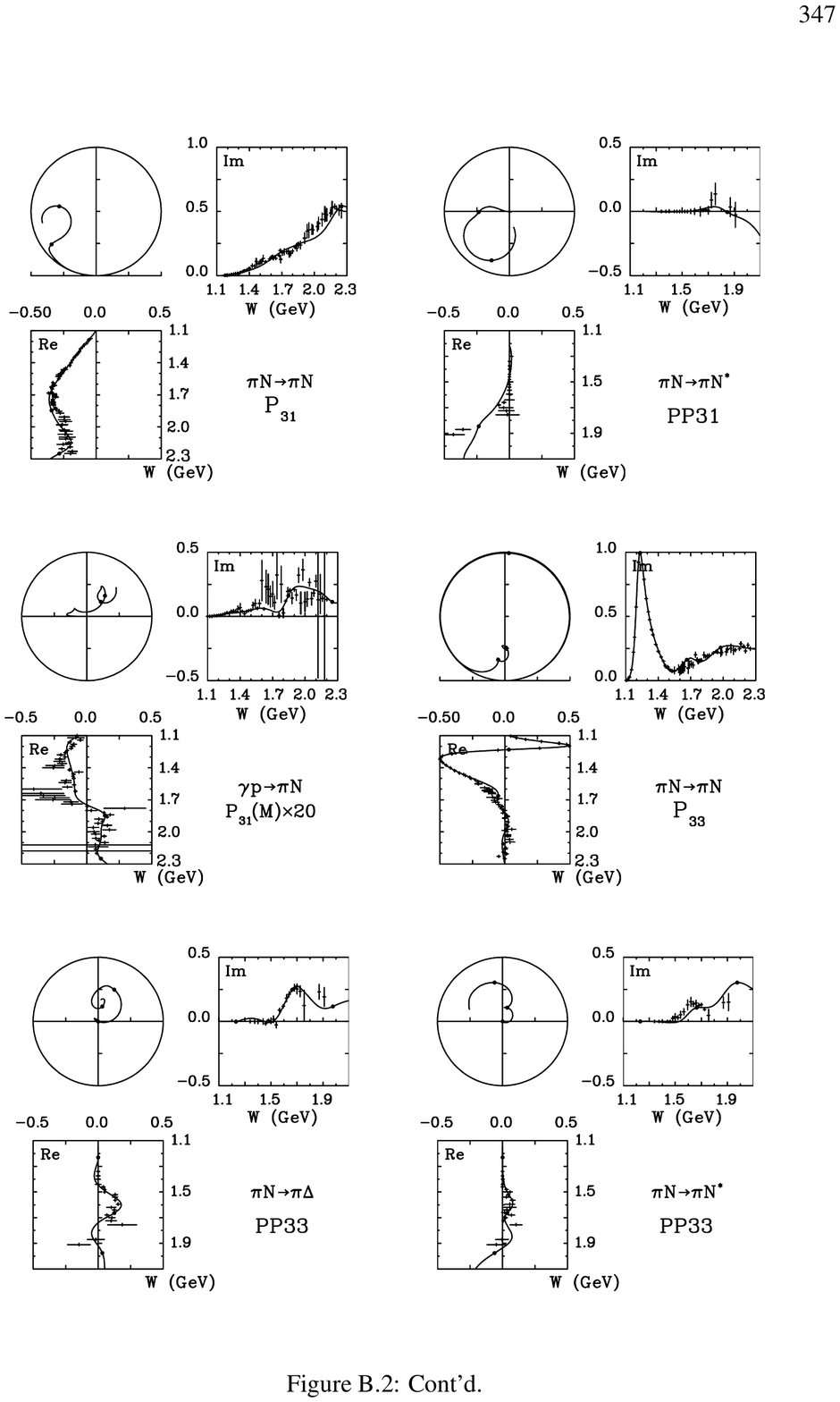}
	\captionof{figure}{Argand diagrams for the $I = 3/2$ amplitudes.}
	\label{Argand22}
\end{minipage}
\newpage
\begin{minipage}[p]{\textwidth}
	\includegraphics[trim={\trimA} {\trimB} {\trimC} {\trimD},clip=true]{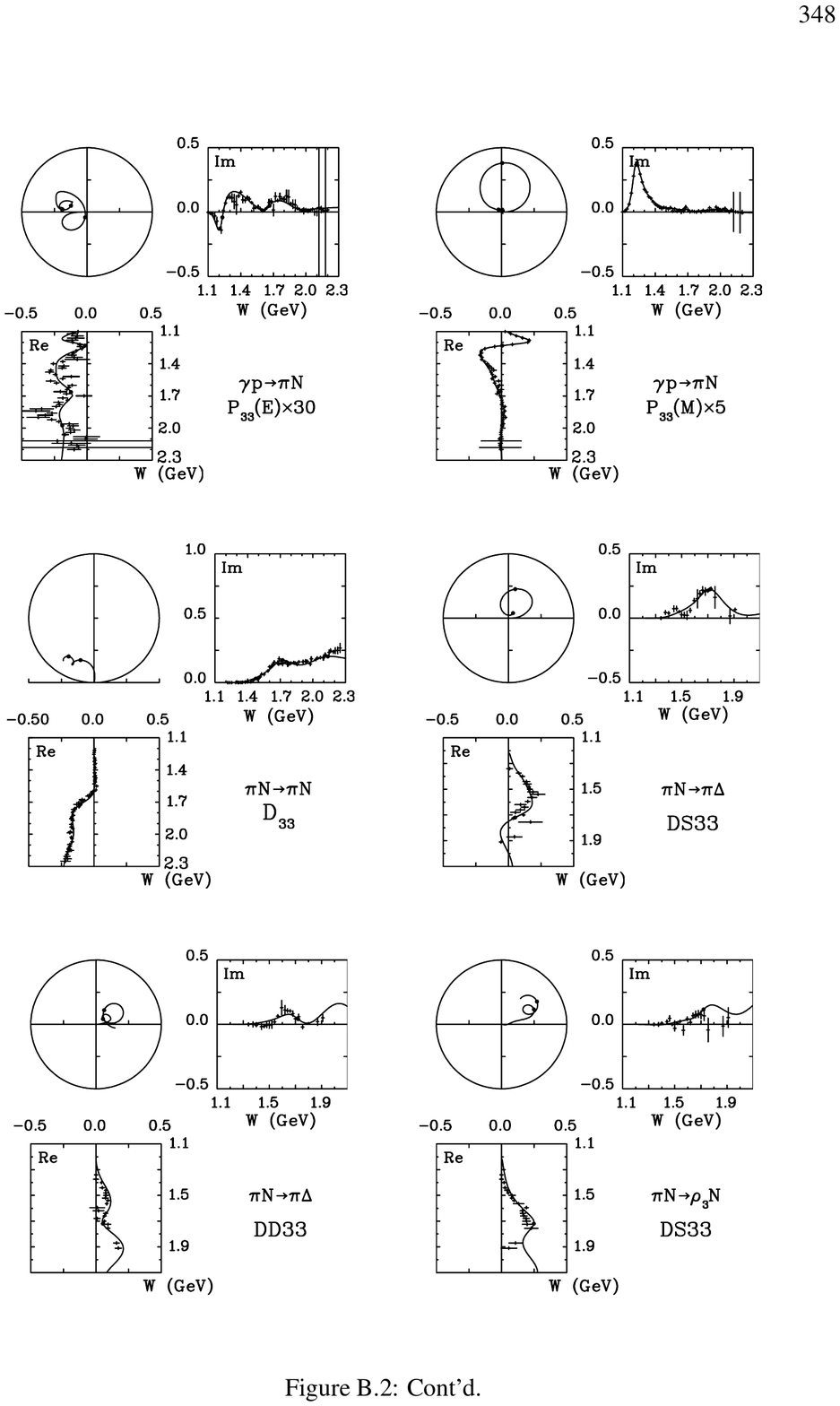}
	\captionof{figure}{Argand diagrams for the $I = 3/2$ amplitudes.}
	\label{Argand23}
\end{minipage}
\newpage
\begin{minipage}[p]{\textwidth}
	\includegraphics[trim={\trimA} {\trimB} {\trimC} {\trimD},clip=true]{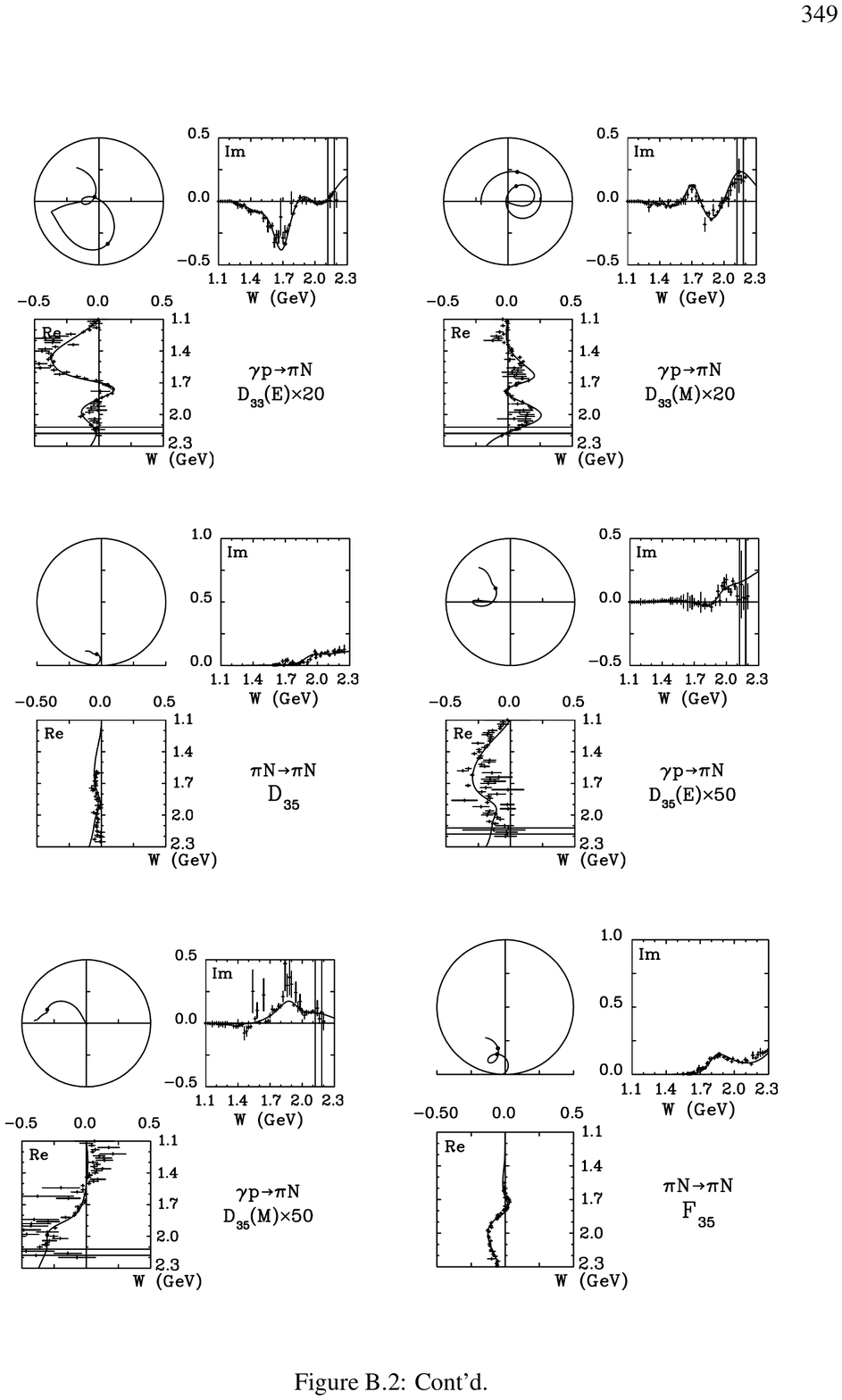}
	\captionof{figure}{Argand diagrams for the $I = 3/2$ amplitudes.}
	\label{Argand24}
\end{minipage}
\newpage
\begin{minipage}[p]{\textwidth}
	\includegraphics[trim={\trimA} {\trimB} {\trimC} {\trimD},clip=true]{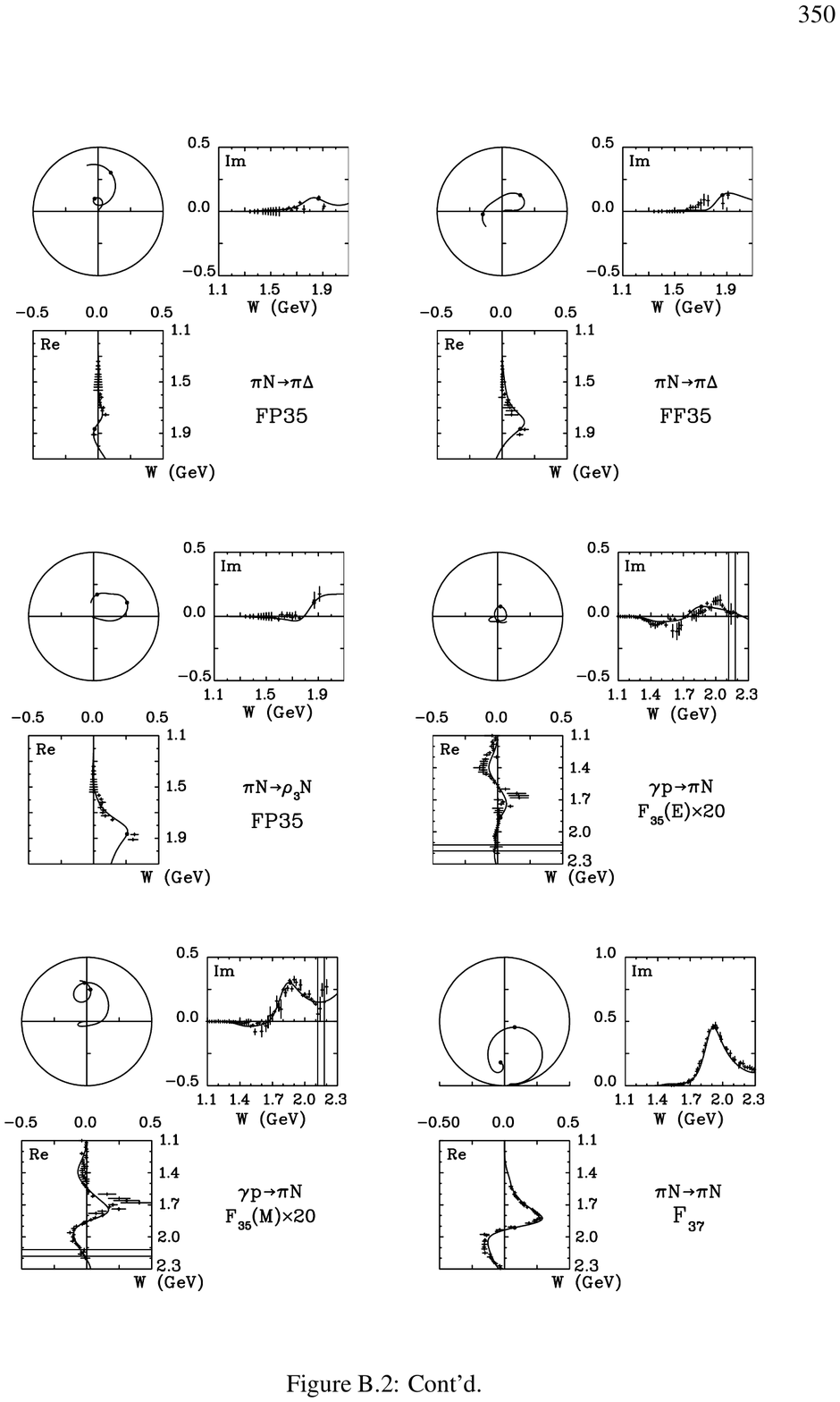}
	\captionof{figure}{Argand diagrams for the $I = 3/2$ amplitudes.}
	\label{Argand25}
\end{minipage}
\newpage
\begin{minipage}[p]{\textwidth}
	\includegraphics[trim={\trimA} {\trimB} {\trimC} {\trimD},clip=true]{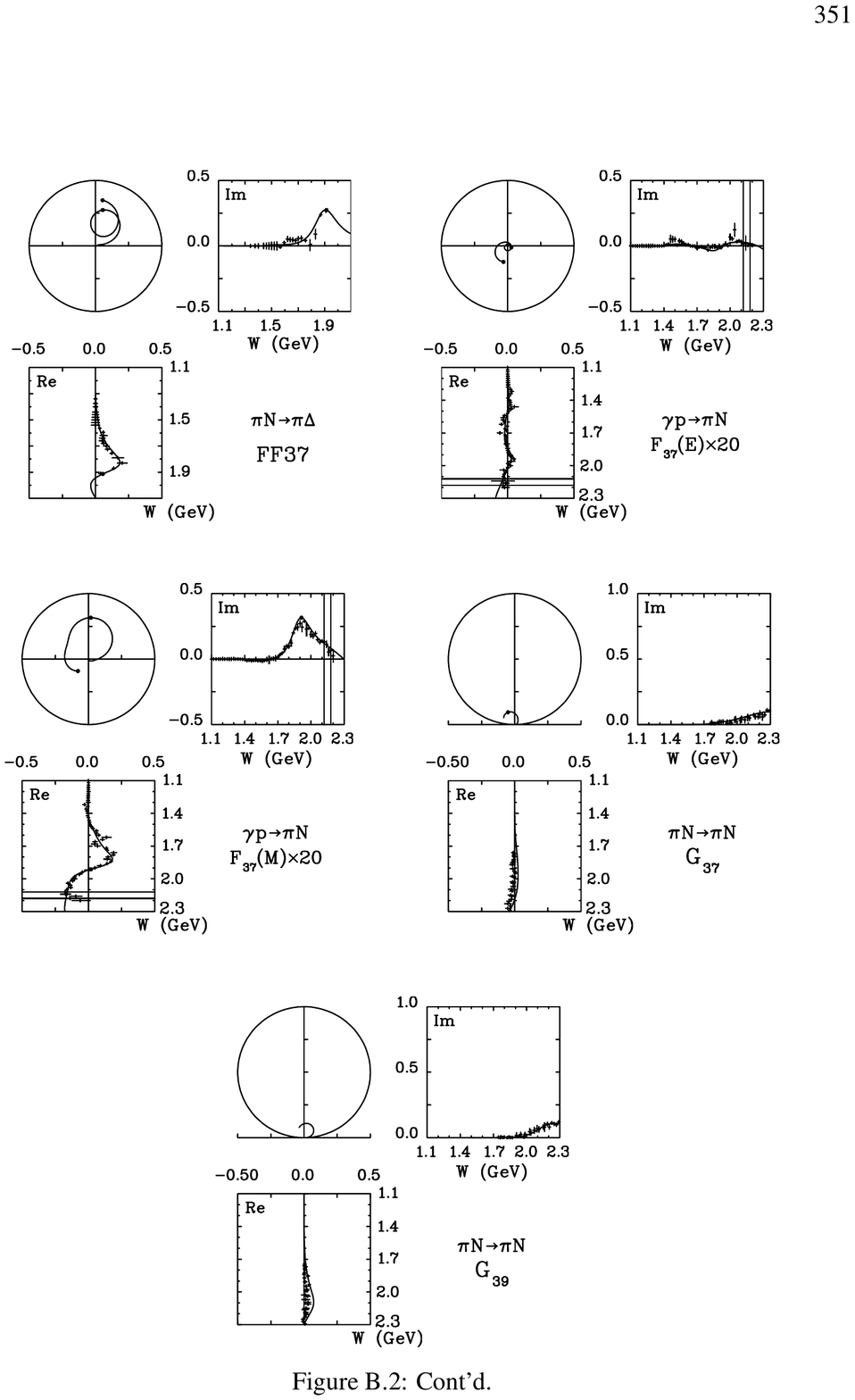}
	\captionof{figure}{Argand diagrams for the $I = 3/2$ amplitudes.}
	\label{Argand26}
\end{minipage}

\bibliography{basename of .bib file}

\end{document}